\documentclass[sigconf]{acmart}

\usepackage{multirow}
\usepackage{tabularx}
\usepackage{subfig}
\usepackage{float}
\usepackage{soul}
\usepackage{enumitem}
\usepackage{colortbl}
\usepackage{xcolor}
\usepackage{bm}
\usepackage[mathcal]{eucal}
\usepackage{stfloats}

\graphicspath{ {figs/} }

\definecolor{lightgray}{gray}{0.6}
\newcommand{\grayline}{\arrayrulecolor{lightgray}\hline\arrayrulecolor{black}}

\usepackage{calc}
\newlength{\DepthReference}
\newlength{\HeightReference}
\newlength{\Width}
\newcommand{\MyColorBox}[2][red]%
{%
    \settowidth{\Width}{#2}%
    \colorbox{#1}%
    {%
        \raisebox{-\DepthReference}%
        {%
                \parbox[b][\HeightReference+\DepthReference][c]{\Width}{\centering#2}%
        }%
    }%
}
\definecolor{redOrganic}{RGB}{255,197,208}
\definecolor{redResponsible}{RGB}{255,197,208}
\definecolor{redCare}{RGB}{255,225,230}
\definecolor{redControl}{RGB}{255,244,246}
\definecolor{verylightgray}{gray}{0.90}
\definecolor{vlightgray}{gray}{0.80}

\makeatletter
\def\gcmidrule{\arrayrulecolor{verylightgray}
    \noalign{\ifnum0=`}\fi
    \@ifnextchar[{\@gcmidrule}{\@gcmidrule[\cmidrulewidth]}}
\def\@gcmidrule[#1]{\@ifnextchar({\@@gcmidrule[#1]}{\@@gcmidrule[#1]()}}
\def\@@gcmidrule[#1](#2)#3{\@@@gcmidrule[#3]{#1}{#2}}
\def\@@@gcmidrule[#1-#2]#3#4{\global\@cmidla#1\relax
    \global\advance\@cmidla\m@ne
    \ifnum\@cmidla>0\global\let\@gtempa\@cmidrulea\else
    \global\let\@gtempa\@cmidruleb\fi
    \global\@cmidlb#2\relax
    \global\advance\@cmidlb-\@cmidla
    \global\@thisrulewidth=#3
    \@setrulekerning{#4}
    \ifnum\@lastruleclass=\z@\vskip \aboverulesep\fi
    \ifnum0=`{\fi}\@gtempa
    \noalign{\ifnum0=`}\fi\futurenonspacelet\@tempa\@xgcmidrule}
\def\@xgcmidrule{%
   \ifx\@tempa\gcmidrule
       \vskip-\@thisrulewidth
       \global\@lastruleclass=\@ne
   \else \ifx\@tempa\morecmidrules
       \vskip \cmidrulesep
       \global\@lastruleclass=\@ne\else
       \vskip \belowrulesep
       \global\@lastruleclass=\z@
   \fi\fi
   \ifnum0=`{\fi}
  \arrayrulecolor{black}}
\makeatother

\newcolumntype{S}{>{\hsize=.12\hsize}X}
\newcolumntype{M}{>{\hsize=.22\hsize}X}
\newcolumntype{B}{>{\hsize=.66\hsize}X}

\newcolumntype{O}{>{\hsize=.1\hsize}X}
\newcolumntype{T}{>{\hsize=.2\hsize}X}
\newcolumntype{E}{>{\hsize=.3\hsize}X}
\newcolumntype{Q}{>{\hsize=.35\hsize}X}
\newcolumntype{f}{>{\hsize=.4\hsize}X}
\newcolumntype{F}{>{\hsize=.5\hsize}X}

\AtBeginDocument{%
  \providecommand\BibTeX{{%
    \normalfont B\kern-0.5em{\scshape i\kern-0.25em b}\kern-0.8em\TeX}}}

\setcopyright{acmcopyright}

\copyrightyear{2022}
\acmYear{2022}
\setcopyright{rightsretained}
\acmConference[CHI `22]{CHI Conference on Human Factors in Computing
Systems}{April 29-May 5, 2022}{New Orleans, LA, USA}
\acmBooktitle{CHI Conference on Human Factors in Computing Systems
(CHI `22), April 29-May 5, 2022, New Orleans, LA, USA}
\acmDOI{10.1145/3491102.3517551}
\acmISBN{978-1-4503-9157-3/22/04}

\begin{document}

\title{Interpretable Directed Diversity: Leveraging Model Explanations for Iterative Crowd Ideation}

\author{Yunlong Wang}
\affiliation{%
  \institution{National University of Singapore}
  \country{Singapore}}
\email{yunlong.wang@nus.edu.sg}

\author{Priyadarshini Venkatesh}
\affiliation{%
 \institution{University College London}
 \city{London}
 \country{UK}}
 \email{zcjtpve@ucl.ac.uk}

\author{Brian Y. Lim}
\affiliation{%
   \institution{National University of Singapore}
  \country{Singapore}}
  \email{brianlim@comp.nus.edu.sg}

\renewcommand{\shortauthors}{Yunlong Wang et al.}
\renewcommand{\shortauthors}{Yunlong Wang, Priyadarshini Venkatesh, and Brian Y. Lim}

\begin{abstract}
  Feedback in creativity support tools can help crowdworkers to improve their ideations. However, current feedback methods require human assessment from facilitators or peers. This is not scalable to large crowds. We propose Interpretable Directed Diversity to automatically predict ideation quality and diversity scores, and provide AI explanations — Attribution, Contrastive Attribution, and Counterfactual Suggestions — to feedback on why ideations were scored (low), and how to get higher scores. These explanations provide multi-faceted feedback as users iteratively improve their ideations. We conducted formative and controlled user studies to understand the usage and usefulness of explanations to improve ideation diversity and quality. Users appreciated that explanation feedback helped focus their efforts and provided directions for improvement. This resulted in explanations improving diversity compared to no feedback or feedback with scores only. Hence, our approach opens opportunities for explainable AI towards scalable and rich feedback for iterative crowd ideation and creativity support tools.
\end{abstract}



\maketitle

\section{Introduction}
Creativity support tools \cite{Frich2019,Shneiderman2007} harness the power of human creativity through large-scale crowdsourcing for tasks, such as text editing \cite{bernstein_soylent:_2010,Clark2018,shah2018building}, iterating designs \cite{Dow2010}, information synthesis \cite{Luther2015}, action planning of health behavior change \cite{agapie_crowdsourcing_2018-3,Hosio2020}, and motivational messaging \cite{Ahmed2019,Cox2021,kocielnik_send_2017,de_vries2016}. Among techniques for creativity support, providing timely and proper feedback is a promising method to boost crowd ideation creativity \cite{bharadwaj2019critter,dow_shepherding_2012,Duque-Estrada2014,Oppenlaender2019,Oppenlaender2020a,yen_listen_2017}. Many feedback methods require human assessment from facilitators or peers, but this limits their ability to scale to large crowds. Employing non-expert crowdworkers can scale more than with experts \cite{Duque-Estrada2014}, but costs can escalate and they cannot provide feedback in real-time. Machine learning can be used to automatically provide feedback by predicting ideation quality and showing examples of high-quality ideations \cite{Peng2020}. However, besides ideation quality, it is also important to increase the diversity and reduce redundancy in submitted ideations \cite{Bjelland2008,KleinMarkandGarcia2015,Riedl2010,Siangliulue2016}. Prior collaborative creativity support methods to avoid redundancy include employing an adaptive task workflows \cite{Yu2011}, constructing a taxonomy of the idea space \cite{Huang2017}, visualizing a concept map of peer ideas \cite{Siangliulue2016}, and directing ideation towards automatically selected diverse prompts \cite{Cox2021}. Current methods to drive diversity provide information from a prior or peer set of ideations, and are not specific to each worker’s ideation. This limits the relevance of feedback \cite{Yang2021}. Therefore, it is important to provide contextualized feedback tailored to the worker’s ideation. In this work, we support both criteria of higher quality and diversity for crowd ideation, specifically targeting the use case of writing motivational messages to promote physical activity.\enlargethispage{8pt}

Although AI can predict scores on ideations, showing scores alone has limited usefulness. Consider receiving feedback in school. To promote learning and improvement, students not only receive graded assignments, but additional feedback to indicate problems in their work, tips or examples for improvement, and opportunities for revision. In the Hattie and Timperley feedback model, these correspond to a score (Feed Up), critical feedback (Feed Back), useful tips or examples (Feed Forward) \cite{Hattie2007}. For ideation writing, we provide each as feedback correspondingly using explainable AI (XAI) with 1) predicted scores, 2) attribution explanation (highlights) of problematic terms, 3) contrastive attributions to provide feedback between revisions, and 4) counterfactual suggestions to provide tips for how to replace problematic terms. We provide feedback across multiple iterations to let workers revise their ideations. Unlike most uses of XAI to improve user understanding and trust of AI \cite{Bansal2021,Lim2009,Tsai2021,Wang2021}, we focus on improving human task performance with human-XAI collaboration \cite{WangDanding2019}. Furthermore, due to the limited use of AI in crowd ideation, the design and effectiveness of XAI for such tasks is an open question. 
%
%
Our contributions  are:
\begin{enumerate}[leftmargin=.7cm]
\item[1.] Interpretable Directed Diversity, an explainable AI approach to predict the quality and diversity of ideations and generate multi-faceted explanatory feedback. This enables scalable, real-time, contextual feedback to improve collective creativity.
\item[2.] Three explanation types --- Attributions, Contrastive Attributions, and Counterfactual Suggestions --- for two criteria, quality and diversity. We implemented the algorithmic approaches in an interactive crowd ideation system.
\end{enumerate}
We evaluated the usage, understandability, and usefulness of the explanations for creative ideation in a formative user study and summative user studies with ideators and validators.
We found that score feedback with attribution, contrastive attribution, and counterfactual suggestion explanations are complementary to improve ideation quality and diversity.
We discuss the generalization of our explainable feedback approach to other domains and explanations.

\section{Related Work}
\subsection{Creative Ideation and Feedback}
Creative ideation involves complex cognitive processes \cite{Amabile1996,BODEN1996267,boden2004creative,Wang2017}. Memory-based explanation models describe how people retrieve information relevant to a cue (prompt) from long-term memory and process it to generate ideas \cite{Adelman1995,Dougherty1999,lubart2001models,Nijstad2006,Nijstad2002}. Ideation-based models \cite{Missier2015} explain how individuals can generate many ideas through complex thinking processes, including analogical reasoning \cite{Green2012,Keller1988,Meheus2000}, problem constraining \cite{Smith2010}, and vertical or lateral thinking \cite{Goel2010}. While users may ideate creatively, their initial ideation can be improved with proper feedback \cite{Dow2010,dow_shepherding_2012,Dow2009}. The efficacy of providing feedback has been much studied in the domains of education \cite{Hattie2007} and management \cite{Ramaprasad1983} to improve learning and decision making. Hattie and Timperley \cite{Hattie2007} argued that effective feedback should answer three questions: where am I going, how am I going, and where to next. Applying goal-setting theory by showing summary feedback \cite{Locke1990,Locke2002}, Carson and Carson \cite{Carson1993} improved individual creativity by manually reviewing ideations and providing feedback on the quantity and creativity of distinct ideas. However, providing feedback often requires intervention from human experts, and this limits their scalability and rapid iteration. In this work, we propose an automated approach for creativity support tools to provide feedback for creative ideation.

{\color{black}Furthermore, we focus on collective creativity from ideations of a crowd of individuals.} While individual creativity is important, society values ideation novelty more. Boden refers to these as P-creativity and H-creativity, respectively \cite{BODEN1996267}. ``An idea is P-creative if it is creative with respect to the mind of the person concerned, even if others have had that idea already. An idea is H-creative if it is P-creative and no other person has had the idea before. H-creativity is more glamorous, but P-creativity is more fundamental" \cite{BODEN1996267}. In crowd ideation, we seek to avoid the wastefulness of redundant ideas, thus we value H-creativity more. Being H-creative requires the ideator to study prior ideations to avoid replicating them. In this work, we support H-creativity by providing automatic prompts and feedback to guide ideators away from prior ideations.

\subsection{Feedback in Creativity Support Tools}
Creativity Support Tools have been widely studied in HCI to help crowdworkers to ideate more effectively and at scale \cite{Frich2019,Frich18}. Effective methods to support crowd ideation include sharing peers’ ideas \cite{Siangliulue2016}, relevant concepts \cite{Bae2020}, and expert guidance \cite{Yang2021}, contextual framing \cite{Oppenlaender2019MUM}, constructing ideation taxonomies \cite{Huang2017}, providing diverse prompts \cite{Cox2021}, and feedback ratings \cite{Peng2020}. We categorize these approaches as manual feedback from people and automated feedback from intelligent systems. Methods using manual feedback investigated who should provide the feedback \cite{dow_shepherding_2012,Oppenlaender2021} and how to coordinate ideators \cite{Duque-Estrada2014,Oppenlaender2020a}. Using Voyant \cite{Duque-Estrada2014}, poster designers had access to a non-expert crowd to receive structural feedback on their designs. Likewise, CrowdUI \cite{Oppenlaender2020a} enables web designers to elicit visual feedback from the website’s community of users. While these methods are adaptable to various domains, they require much human labor to prepare feedback and are difficult to scale. 

In contrast, recent works have developed automatic systems to generate feedback for mind mapping \cite{Bae2020}, story writing \cite{Clark2018}, metaphor creation \cite{Gero2019}, and writing supportive comments for online mental health community \cite{Peng2020}. The feedback provided in these works were limited to independent examples from peers or machine-suggested words or sentences, but not contextual to the ideations. MepsBot \cite{Peng2020} predicts scores for informational and emotional support, which is specific to the submitted ideations. However, the justification for such scores is unclear to users, hence we provide AI explanations in this work for more actionable feedback.

Finally, prior works focus on individual creativity, so ideations from multiple ideators may be redundant. Directed Diversity \cite{Cox2021} addressed this problem by automatically selecting diverse prompts for ideation, but did not provide feedback after ideation. In this work, we address the difficult problem of providing contextual feedback for collective ideation, where each ideator cannot see all prior messages. We achieve this by providing model explanations towards a higher automatically-calculated diversity score. 

\subsection{Explanations in Intelligent Systems}
Explainable AI (XAI) techniques have drawn much attention as intelligent systems become more complex, requiring transparency and accountability \cite{Abdul2018,WangDanding2019}. Prior research has shown that XAI can increase user trust and understanding of AI systems \cite{Bansal2021,Wang2021}, and explored designing proper XAI to improve human-AI collaboration, e.g., for computer-aided translation \cite{Coppers2018}, playing Chess \cite{Das2020}, and music creation \cite{Louie2020}. Though many studies have sought to identify optimal explanation types in specific tasks \cite{Lim2009,Tsai2021}, some found that providing a variety of explanations can have stronger benefits \cite{Anderson2020} or support various usage strategies \cite{Lim2011,Lim2013}. While prior studies have found that XAI can improve system transparency, fairness, and acceptance, we explore how XAI can improve user creativity and employed multiple explanations to stimulate creativity.

\section{Technical Approach}
\begin{figure}
  \centering
  \includegraphics[width=1\linewidth]{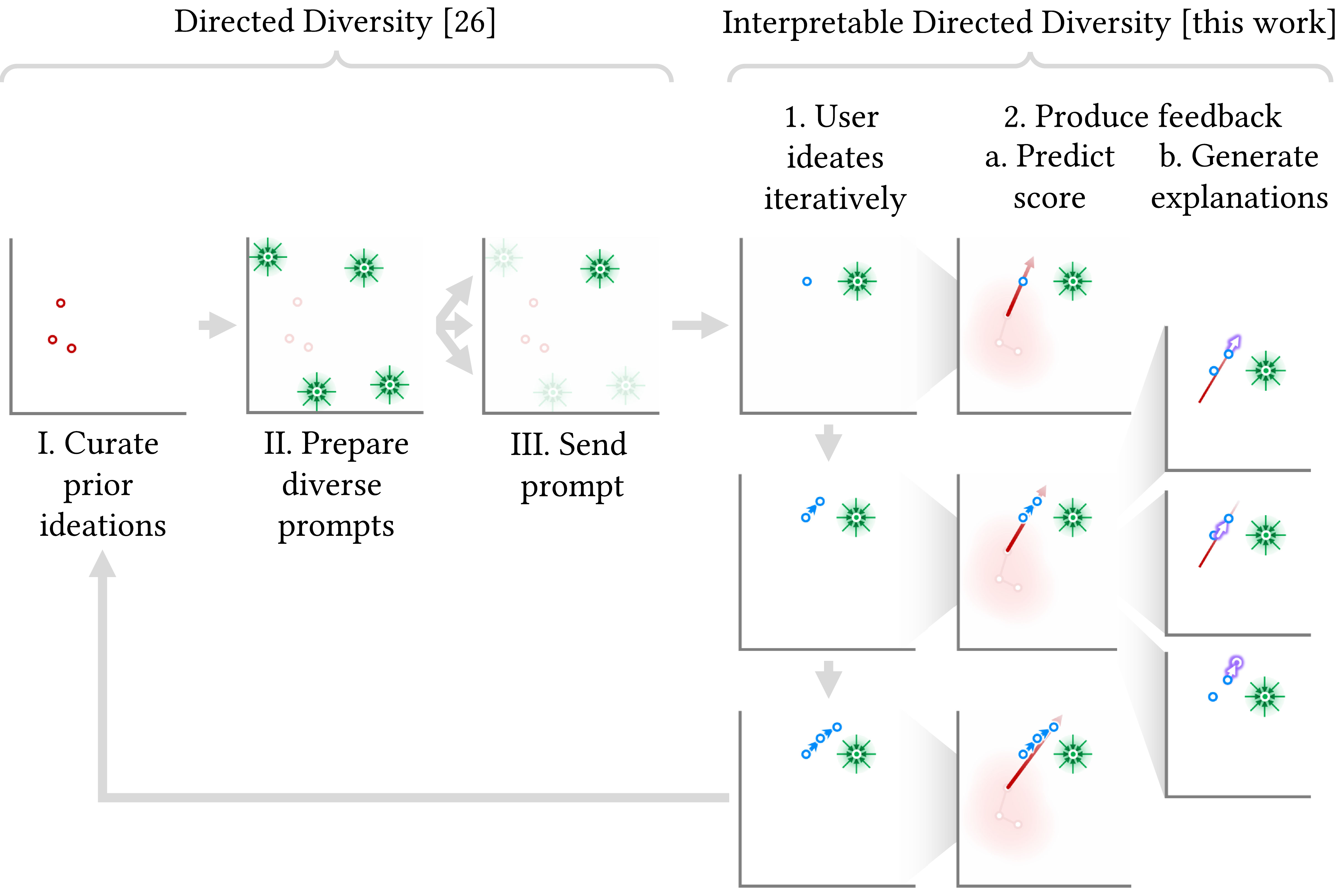}
  \caption{Overall technical pipeline to curate collected ideations, prepare diversely selected prompts, send the prompts, and provide feedback (as scores and three explanation types) for iterative ideation. Dots represent prompts (green) and ideations (red for prior, blue for iterations of new ideation) as points in a 2D vector space of ideas. A prompt will attract ideations towards it, though influence will depend on prior iterations and explanation type.}
  \Description{Abstract scatter plots showing the overall technical pipeline to curate collected ideations, prepare diversely selected prompts, send the prompts, and provide feedback. Ideations and prompts are shown as dots and the distance between dots represents semantic difference.}
  \label{fig1}
\end{figure}

We aim to direct ideators towards improved ideations by providing automatic feedback and explanations. For each prompted ideation, ideators can learn from the feedback to iteratively improve their ideations. Figure \ref{fig1} shows the iterative process of Interpretable Directed Diversity. It involves two high-level phases to prepare and provide prompts (Steps I, II, III), and analyze the ideation and provide feedback iteratively (Steps 1, 2a, 2b).  Phase 1 follows the same steps as in Directed Diversity \cite{Cox2021} to I) curate prior ideations, II) generate diverse and non-redundant prompts, and III) send these prompts to ideators. Phase 2 is the main focus of this paper, extends Directed Diversity \cite{Cox2021} and involves, for each ideation, 1) analyzing the written ideation, 2) showing feedback based on a) prediction scores and b) explanations; the analysis and feedback are repeated as the ideator iteratively revises each ideation. The technical contribution of this work is the predictive and explanatory feedback after the prompting and ideation in Directed Diversity \cite{Cox2021}.

\subsection{Feedback Score from Prediction Models}
The base feedback provides assessment scores of the ideation. Currently, we predict two scores using a machine learning model and a heuristic model. The first goal for ideation is typically to ensure high-quality ideas, hence, we predict a quality score $s_q$. However, to mitigate redundancy in crowd ideation \cite{Cox2021}, another goal is to increase diversity, which we predict with a diversity score $s_d$. Hence, we integrate both scores in our modeling.\enlargethispage{10pt}

\subsubsection{Quality Score Prediction}
We predicted quality using a machine learning model, since it is rated by people. We trained the prediction model based on data collected by Cox et al. \cite{Cox2021}. The training dataset consists of 815 ideation messages $\bm{x} \in X$ written by 250 ideators and rated by 150 validators on quality (motivating). Ideators wrote messages to encourage physical activity and were shown various prompts (None, Random, Directed Diversity). Each message was rated by M=4.6 validators on a 7-point Likert scale (–3 very demotivating to +3 very motivating). We model the prediction problem as a binary classification instead of regression on 7 intervals to simplify it, because subjective ratings are noisy and suffer from individual variance.  Furthermore, we binarized the rating of each message to whether it was highly motivating (above Median = 1.17) to create a balanced dataset (equal number of positive and negative classes) for better model training. While this approach includes borderline ratings that may train an inaccurate decision boundary, it benefits from training with a larger dataset. See Table \ref{table4} in Appendix A.1 for example training data. Similar  to \cite{Cox2021}, we computed the Universal Sentence Encoder (USE) \cite{Cer2018} embedding $\bm{z}$ for each ideation to represent its idea as a vector. Using this embedding and normalized message length as input features, we trained a two-layer fully-connected neural network with sigmoid activation in the final layer to predict quality. The model was accurate with AUC = 0.717 from a 5-fold cross-validation. We determine the quality score $s_q$ as the prediction confidence (0-100\%). Note that this represents the probability of the message being highly motivating, rather than the estimated rating of a validator in \cite{Cox2021}. 

\subsubsection{Diversity Score Calculation}
While \cite{Cox2021} had also collected validator ratings on perceived diversity, those measures relate to pairs or groups of messages rather than individual ones. We require a single diversity label for each message to provide direct feedback during ideation, hence we use a heuristic calculation. We determined the diversity score $s_d$ by adding the new ideation $\bm{x}$ to the prior ideations $X$, and calculating the increase in the Ideation Dispersion (MST Sum of Edge Weights) metric as defined in \cite{Cox2021} from the prior collection to the new collection with the new ideation. For evaluation (later section), we initialized with 50 prior ideations from messages collected by Cox et al. \cite{Cox2021}.

\subsection{\hbox{Feedback Explanations from Explainable AI}}
Inspired by Miller \cite{Miller2019}, we propose actionable explanations to increase ideation quality and diversity --- attribution, contrastive attribution, and counterfactual suggestions. These provide multi-faceted feedback for ideators to understand issues and opportunities to ideate better. Our explanation techniques are agnostic to the prediction models and can generalize to multiple scores. For brevity, we combine the notation for the quality and diversity scores as vector $\bm{s}=(s_q,s_d)^\top$. Next, we describe each explanation type in detail.\enlargethispage{10pt}

\subsubsection{Attribution Explanation}
Attribution explanations answer the question ``Why P". They indicate which features are influential in a prediction. For ideated messages, we treat each word as a feature. Each attribution may support (positive value) or undermine (negative) the prediction. The larger the magnitude, the stronger the influence. Since the goal is to increase the score, the explanation should explain which features most hinder a higher score, and ideators should focus on words with the most negative attribution. There are several methods for calculating feature attributions, such as calculating their gradients \cite{Sundararajan2017}, backpropagated relevance \cite{Binder2016}, Shapley values \cite{Lundberg2017}, or approximating linear slopes \cite{Ribeiro2016}. However, these methods are computationally expensive, since they involve iterating through many parameters, features, or neighboring instances. This is infeasible for our application, since we need fast calculations for live feedback and cannot pre-compute explanations. Instead, we use a simpler sensitivity analysis based on ablation \cite{Richardson2006,Tullio2007}. Our approach to calculating attributions are as follows:
\begin{enumerate}[leftmargin=.5cm]
\item[1.] Define the ideation message as $\bm{x}$ with the $r$th word as $x_r$
\item[2.] Ablate (remove) the \textit{r}th feature $x_r$ from the dataset
\item[3.] Calculate the score of the new simulated ideation $\bm{s}(\bm{x}\setminus{\{x_r\}})$
\item[4.] Calculate the feature attribution $w_r$ as the decrease in the predicted score $\bm{s}$, i.e., $w_r=-(\bm{s}(\bm{x})-\bm{s}(\bm{x}\setminus{\{x_r\} }))$
\item[5.] Shift all attributions to be negative to emphasize the most important features to change with largest magnitude, i.e., $w_r \rightarrow w_r-\min_{\rho\in R}(w_\rho)$. This makes the attributions more actionable.
\end{enumerate}

\begin{figure}
  \centering
  \includegraphics[width=1\linewidth]{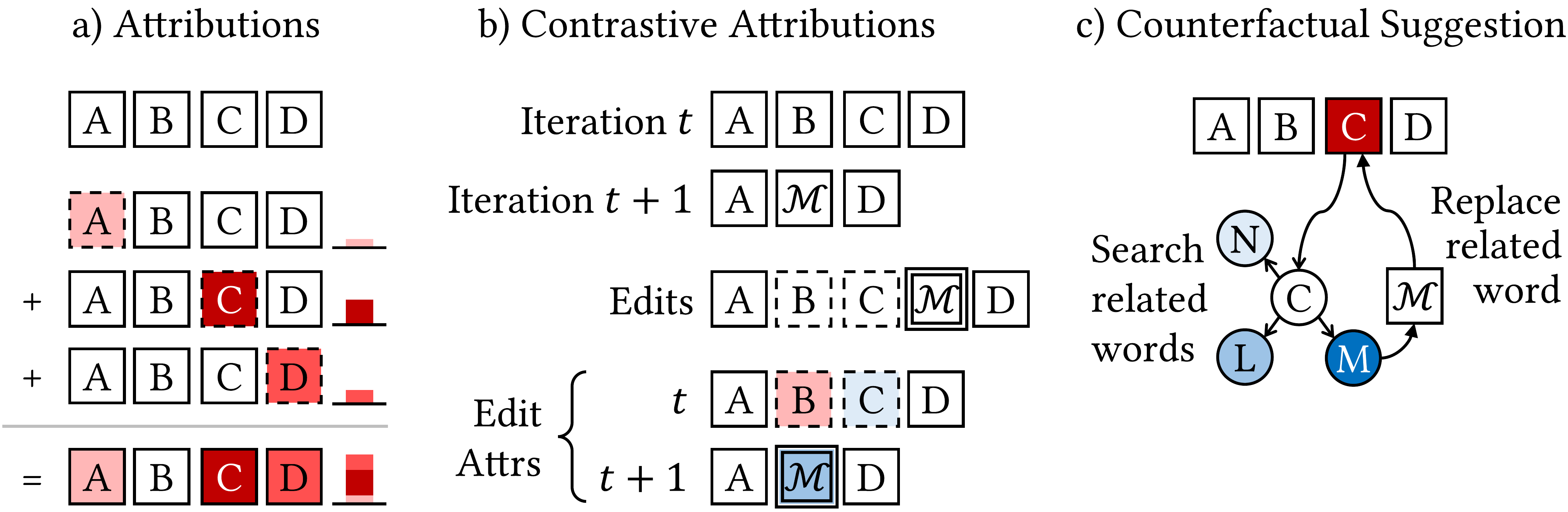}
  \vspace{-0.3cm}
  \caption{Conceptual approach for explanations. Each square represents a word, each letter for a different word. A word can be existing (solid line), inserted (double-line), or deleted (dotted). Colors indicate attribution: red = word that should be changed to increase score, blue = good word that increased or could increase the score, darker color = higher score magnitude. a) For an ideation (top row), the attribution towards a word is based on the score decrease (right stacked bar chart) when it is deleted from the ideation (e.g., A, C, D). Their sum is the total Attributions. b) Contrastive attributions compare an ideation $t+1$ (second row) with its previous iteration $t$ (first). B and C were deleted, while $\mathcal{M}$ was inserted. Attributions for these edits show that deleting C and adding $\mathcal{M}$ were beneficial, while deleting B was detrimental. c) Counterfactual Suggestions involve searching for related words from a knowledge graph (lower left) and calculating their Attribution towards increasing the scoring. Using the suggested word M as inspiration, the ideator may replace C with $\mathcal{M}$.}
  \Description{Abstract figure showing different types of explanations. Each word in an ideation is shown as a square. Colors of squares are used to indicate the relative values in explantions.}
  \label{fig2}
\end{figure}

Figure \ref{fig2}a illustrates this algorithm. This approach can be applied to the Diversity and Quality scores, since it is agnostic to how the score is calculated (can be machine learning or heuristic). See the red highlights (negative attributions) of the feedback user interface in Figure \ref{fig6} for an example of attributions explanation.

\subsubsection{Contrastive Attribution Explanation}
\begin{figure}
  \centering
  \includegraphics[width=1\linewidth]{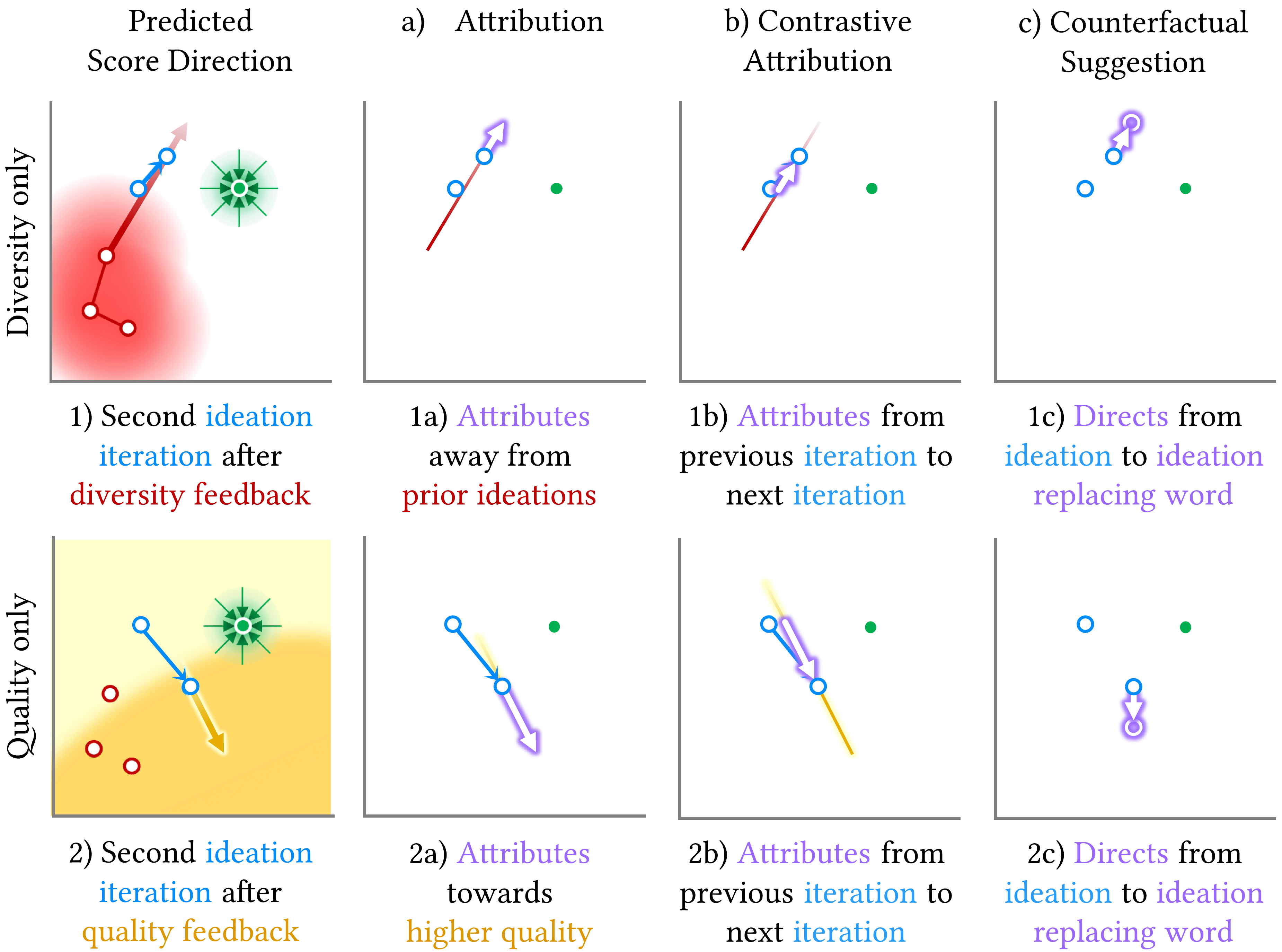}
  \vspace{-0.3cm}
  \caption{Conceptual effect of providing different types of explanations for ideation feedback with the intent to increase the score for diversity (top row) or quality (bottom row) separately. Dots represent prompts (green) and ideations (red for prior, blue for iterations of new ideation) as points in a 2D vector space of ideas. 1) Darker red regions indicate locations in the vector space that are dense with prior ideations that new ideations should avoid increasing diversity. 2) Yellow areas indicate a nonlinear decision surface with the change in color representing a sharp decision boundary. 1,2) Purple arrows indicate the directional influence of the feedback. a) Attribution explanations direct ideation towards a higher score. b) Contrastive attribution explanations convey the difference between iterations along the direction of a higher score. c) Counterfactual suggestion explanations direct towards an alternative, similar ideation with a higher score.}
  \Description{Abstract scatter plots showing the hypothetical effect of different feedback on ideation diversity and quality. Ideations and prompts are shown as dots and the effects of feedback are shown as arrows.}
  \label{fig3}
\end{figure}

Contrastive explanations answer the question ``Why P and not Q". The contrastive attribution explanation extends the attribution explanation to focus on specific differences between two classification labels, and we extend their use to contrast between two iterations, $\bm{x}^{(t_1 )}$ and $\bm{x}^{(t_2 )}$, of an ideation. This identifies the edit differences in word attributions between them, which are either insertions or deletions. For simplicity, we do not distinguish word order. We generate contrastive explanations for ideation iteration with these steps:\enlargethispage{10pt}
\begin{itemize}[leftmargin=.5cm]
\item[1.] For each inserted word $x_{r\in R_{ins}}$, calculate attribution as $w_{r\in R_{ins}}^{(t_2)}=(\bm{s}(\bm{x}^{(t_2)})-\bm{s}(\bm{x}^{(t_2)}\setminus \{x_{r\in R_{ins}}\}))$
\item[2.] For each deleted word $x_{r\in R_{del}}$, calculate attribution in reverse, i.e., add the word to the later iteration $x^{(t_2)}$ and compute the decrease in score, as $w_{r\in R_{del}}^{(t_2)}=-(\bm{s}(\bm{x}^{(t_2)})-\bm{s}(\bm{x}^{(t_2)}\cup \{x_{r\in R_{del}}\}))$
\item[3.] Calculate the change in scores, i.e., $\Delta s = \bm{s}(\bm{x}^{(t_2)})-\bm{s}(\bm{x}^{(t_1 )})$
\item[4.] Calibrate total attributions to match the change in scores, i.e.,
    \begin{itemize}
         \item[1.] Min-max normalize all attributions to between 0 and 1
         \item[2.] Shift the attributions such that $\sum_{r\in { \{R_{ins} \cup R_{des} \} } } w_r^{(t_2)} =\Delta s$
      \end{itemize}
\end{itemize}
Figure \ref{fig2}b illustrates this algorithm. See the red (negative attributions) and blue (positive) highlights of the feedback user interface in Figure \ref{fig7} for an example of contrastive attributions explanation.

\subsubsection{Counterfactual Suggestion Explanation}
Counterfactual explanations answer ``How to change inputs to get Q instead of P". For ideation, this determines how an ideation could be edited to increase its score. We propose to substitute words with alternative suggestions searched using the ConceptNet knowledge graph v5.8 \cite{Speer2017} based on various semantic relationships. This will reduce the cognitive load, mitigate the ideator’s lack of experience with recalling related terms, and stimulate more ideas \cite{Bae2020}. In an ideation $\bm{x}$, for each important feature word $x_r$ with negative attribution,\enlargethispage{10pt}
\begin{itemize}[leftmargin=.5cm]
    \item[1.] Search for related words $x_{\rho_r}$ using the knowledge graph
       \begin{itemize}
           \item[1.] Exclude feature words $x_r$ with too few (<10) related words 
           \item[2.] Exclude relations\footnote{Excluded: Synonym, Antonym, DerivedFrom, SymbolOf, DefinedAs,MannerOf, EtymologicallyRelatedTo, EtymologicallyDerivedFrom, ExternalURL. 
           } that are less actionable
       \end{itemize}
    \item[2.] For each related word $x_{\rho_r}$,
       \begin{itemize}
           \item[1.] Substitute feature word $x_r$ with its related word $x_{\rho_r}$ into the ideation message $\bm{x}$
           \item[2.] Filter word $x_{\rho_r}$ for relevance
               \begin{itemize}
                   \item[1.] Compute USE \cite{Cer2018} embedding vector $\bm{z}_r$ for the word $x_{\rho_r}$
                   \item[2.] Compute mean pairwise distance $\bar{d}_{\rho_r}$ to prior ideations
                   \item[3.] Exclude word if out-of-scope, like in \cite{Cox2021}, i.e., $\bar{d}_{\rho_r}>\delta$
               \end{itemize}
           \item[3.] Calculate attribution $w_{\rho_r}$ due to deleting $x_r$ and inserting $x_{\rho_r}$, i.e., $w_{\rho_r }=-(\bm{s}(\bm{x})-\bm{s}(\bm{x} \cup \{x_r\} \setminus \{x_{\rho_r})\}))$
           \item[4.] Include related word with large attribution, i.e., $w_{\rho_r}>\omega$
\end{itemize}
\end{itemize}
Figure \ref{fig2}c illustrates this algorithm and Figure \ref{fig8} shows a demo. We limited Counterfactual suggestions for words with negative attribution. The approach cannot replace phrases or whole sentences.

\subsubsection{Summary and Hypothesized Effects of Explanations}
We proposed three actionable explanations for improving ideation scores. Specifically, we a) shifted Attribution explanations to frame influence by which words are best to improve the ideation; b) framed Contrastive Attributions by what changes were successful or detrimental; and c) streamlined Counterfactual Suggestions to recommend non-distant words that are estimated to improve scores.

These explanations aim to direct ideators towards higher scores, but have slight differences in direction (Figure \ref{fig3}). The Attributions explanation (Figure \ref{fig3}a) points in the direction of increasing score, but may be prone to some error due to the approximations in the ablation technique (e.g., not calculating Shapley values \cite{Lundberg2017}). The Contrastive Attributions explanation (Figure \ref{fig3}b) describes how the difference between two iterations works towards or away from increasing score; this is equivalent to resolving the blue arrow vector along the red/yellow line vector. It is also subject to approximation errors like Attributions. The Counterfactual Suggestion explanation (Figure \ref{fig3}c) directs the ideator towards a hypothetical ideation with the substituted word(s)
, which may not be the most direct towards
increasing score. Finally, all three explanation types do not necessarily direct the ideation straight towards the original prompt, but this can also provide opportunities for diversification.
\enlargethispage{12pt}

\section{Prompt and Feedback Interfaces}
We describe the ideation interface variants and how to use them.

\subsection{Ideation Task and Prompt Interface}
\begin{figure}
  \centering
  \includegraphics[width=1\linewidth]{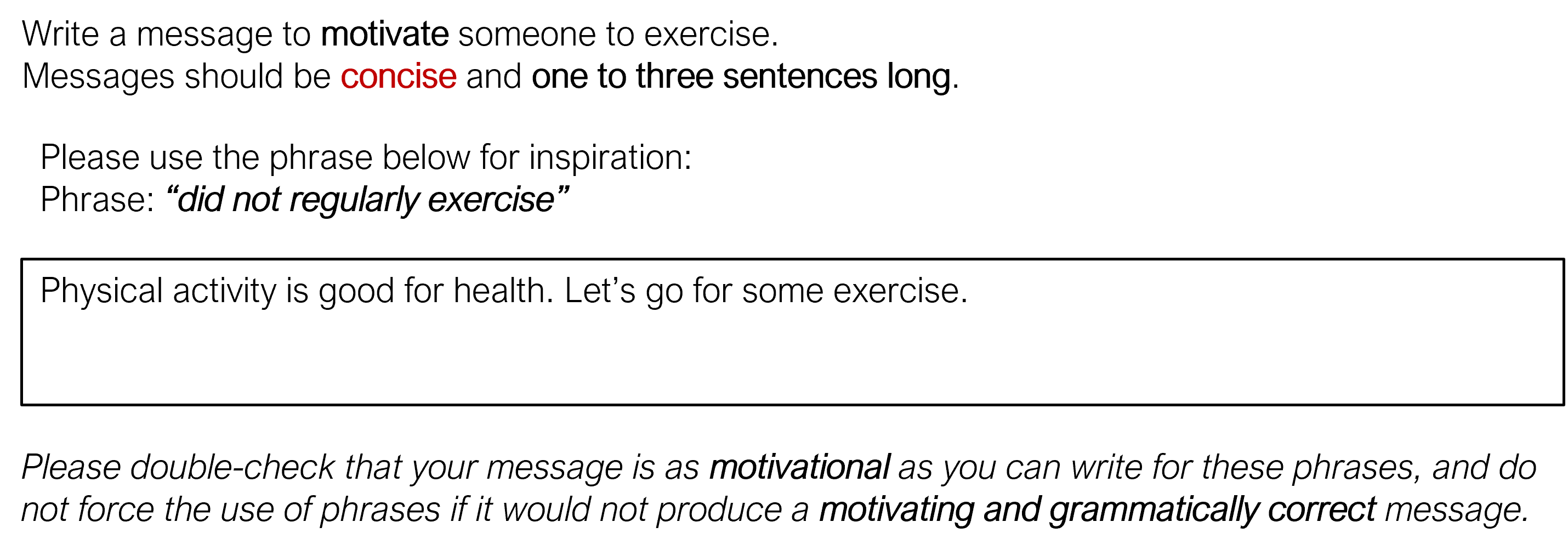}
  \vspace{-0.6cm}
  \caption{Prompt UI for ideation.}
  \Description{Screenshot of Prompt UI for ideation. The phrase "did not regularly exercise" is shown to inspire an ideator to write a message to motivate exercise. The written message is "Physical activity is good for health. Let's go for some exercise."}
  \label{fig4}
\end{figure}

Consider the ideator’s task to write motivational messages to encourage exercise and physical activity. For each ideation, she is prompted with a phrase and asked to write a message inspired by the phrase (Figure \ref{fig4}). She does not need to use the words or concepts if she finds them too irrelevant or awkward. After the first iteration, the participant is shown feedback and asked to revise the ideation, up to two times. For each iteration, she is reminded of the fixed original prompt and shown feedback based on her writing.

\newcolumntype{Y}{>{\centering\arraybackslash}X}
\begin{table}
  \caption{Feedback Conditions in user studies. We investigated 6 variants of the Feedback Interface (columns), formed from combinations of four Feedback Features (rows). Feedback Interface is the main independent variable in our experiments.}
  \vspace{-0.3cm}
  \label{table1}
  \small
  \begin{tabularx}{.48\textwidth}{r c c YYYY}
    \toprule
     & \multicolumn{6}{c}{\textbf{Feedback Interface}} \\
    \cmidrule{2-7}
    \textbf{Feedback Feature} & N & S & SA & SAX & SAC & SAXC \\
    \midrule
    Score Prediction (S) &  & \checkmark & \checkmark & \checkmark & \checkmark & \checkmark\\
    \grayline
    Attributions (A) &  &  & \checkmark & \checkmark &  \checkmark & \checkmark\\
    Contrastive Attributions (X) &   &   &   & \checkmark  &   & \checkmark \\
    Counterfactual Suggestions (C) &   &   &   &   &  \checkmark & \checkmark \\
    \bottomrule
\end{tabularx}
\end{table}

\begin{figure}
  \centering
  \includegraphics[width=1\linewidth]{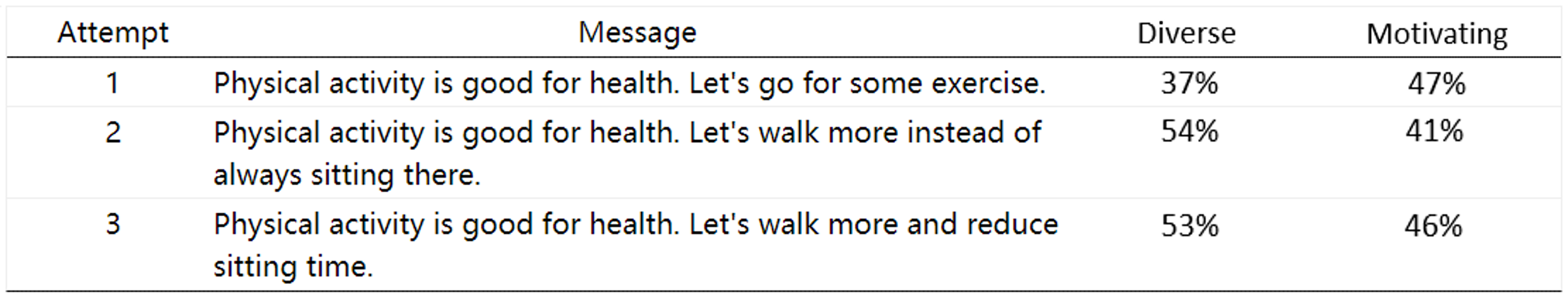}
  \vspace{-0.6cm}
  \caption{User interface with Score feedback (S) showing Diversity and Quality scores for each iteration.}
  \Description{Screenshot of UI with Diversity and Quality scores for each iteration.}
  \label{fig5}
\end{figure}

\begin{figure}
  \centering
  \includegraphics[width=1\linewidth]{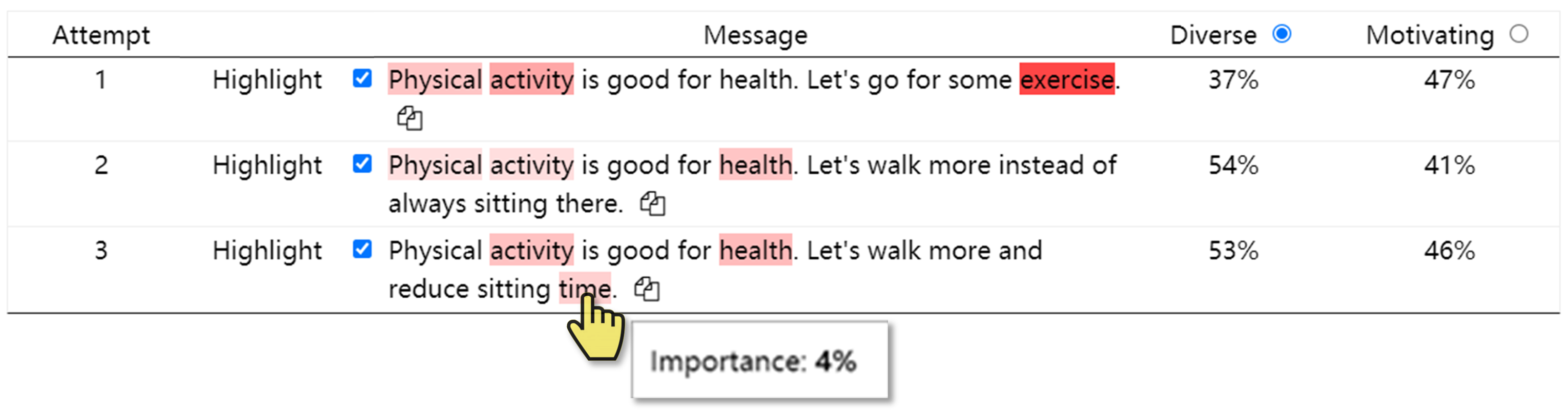}
  \vspace{-0.75cm}
  \caption{Score + Attribution (SA) feedback showing the most attributed words for the selected score (Diversity, in this case). Darker red indicates higher importance. Hovering over a highlighted word shows the attribution sub-score.}
  \Description{Screenshot of UI with Attribution explanations, besides Diversity and Quality scores for each iteration. For each ideation, three words are highlighted to be prioritized for revision.}
  \label{fig6}
\end{figure}

\begin{figure}
  \centering
  \includegraphics[width=1\linewidth]{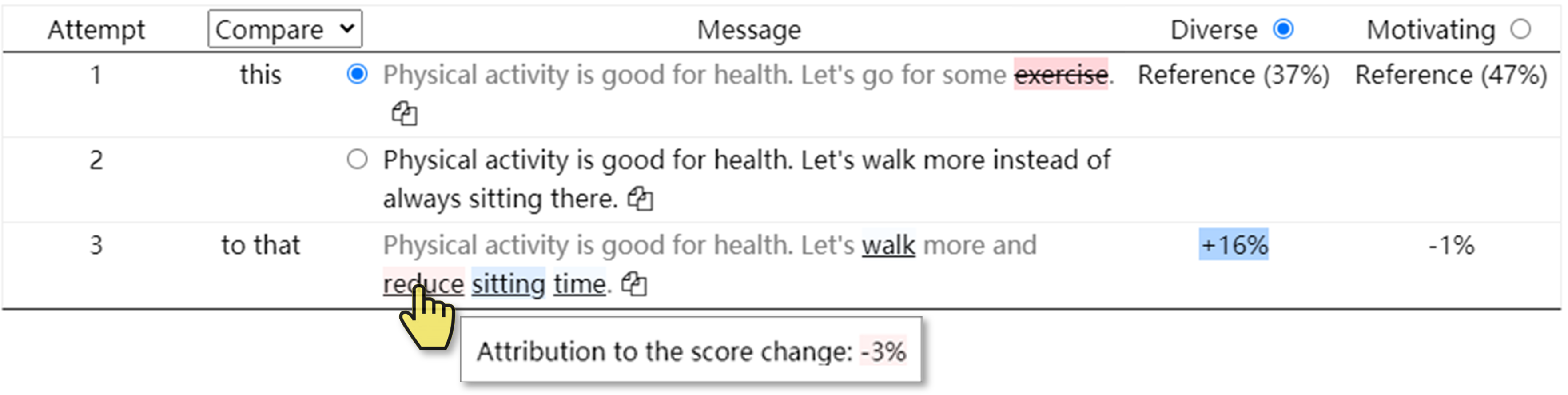}
  \vspace{-0.75cm}
  \caption{SA + Contrastive Attribution explanations (SAX) feedback comparing Attempts 1 and 3. Underlined words were inserted in Attempt 3 compared to 1, and struck-out words were deleted from Attempt 1 to 3. Inserting words ``walk", ``sitting" and ``time", and deleting ``exercise" increased Diversity score, but inserting ``reduce" decreased the score.}
  \Description{Screenshot of UI with Contrastive Attribution explanations, besides Diversity and Quality scores for each iteration. The edited words between two iterations are highlighted with different colors to show their effect on the change of ideation score.}
  \label{fig7}
\end{figure}

\begin{figure}
  \centering
  \includegraphics[width=1\linewidth]{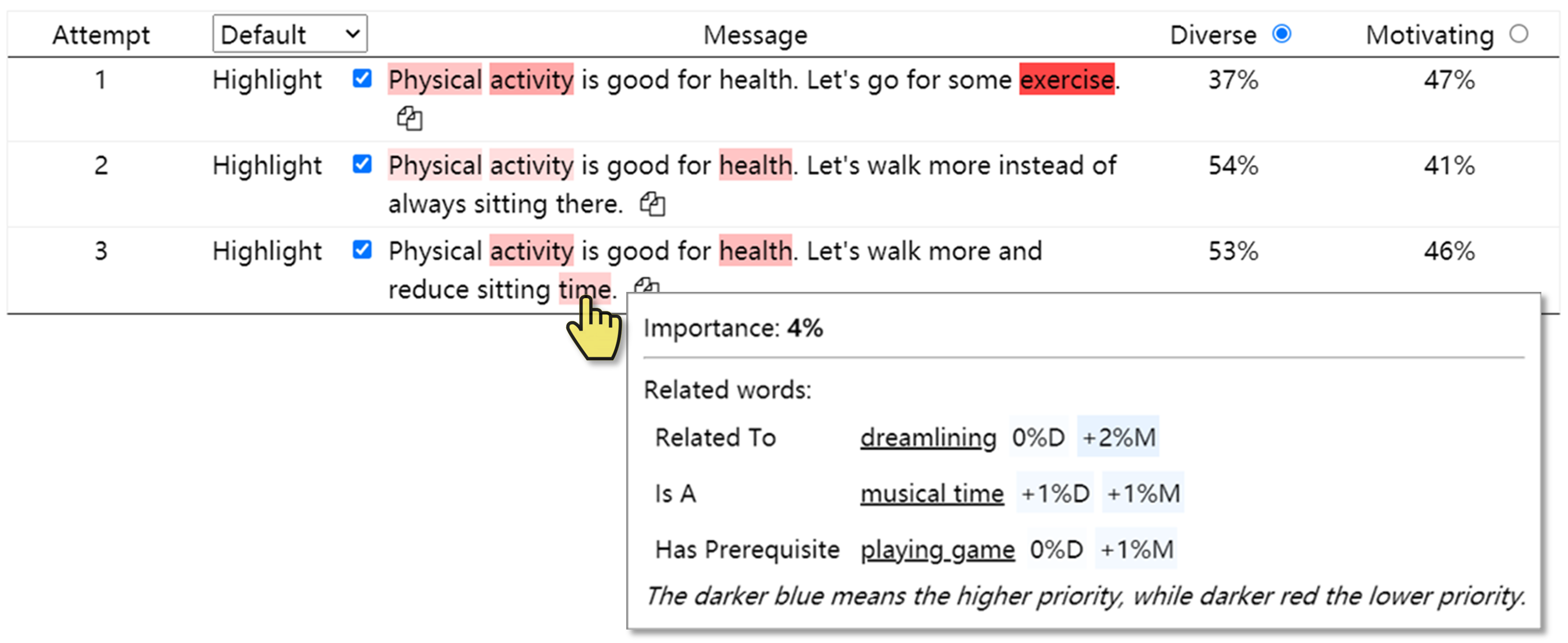}
  \vspace{-0.75cm}
  \caption{SAX + Counterfactual Suggestion explanations (SAXC) feedback. 
  Hovering a highlighted word will show a pop-up with attribution importance, and suggested replacement words. Potential score increases indicated with blue highlights; darker blue for higher increase. This suggests changing ``time" to ``musical time" to increase diversity by +1\%, but to "dreamlining" to increase quality by +2\%.\vspace*{-12pt}}
  \Description{Screenshot of UI with Counterfactual Suggestion explanations, besides Attribution explanations and scores for each iteration. Three suggested words are shown when hovering a highlighted word.}
  \label{fig8}
\end{figure}

\subsection{Feedback Features and Interfaces}
We investigated 6 variants (see Table \ref{table1}) of the Feedback Interface that had different combinations of whether to show explanations and which type (Feedback Features). Note that since all XAI-based feedback explain the Score, they cannot omit showing it. Furthermore, since attribution is foundational to contrastive attribution and selecting words for counterfactual suggestions, all XAI-based feedback interfaces include Attribution explanations.

The first baseline interface with \textbf{No feedback (N)} (Feedback Condition = N, Figure \ref{fig5}, left two columns only) only shows the previous messages that the ideator wrote in separate rows in a table. The ideator is only asked to write the next iteration without any other information. The second baseline interface, adds \textbf{Scores (S)} on the right columns of the feedback table (Figure \ref{fig5}). Ideators can see two scores for motivating-ness (quality) and diversity increase. The quality score shows confidence \% of the model when predicting high quality. The diversity score was normalized with 100\% for the maximum possible angular distance ($\pi$ on the unit hypersphere).

We next describe the XAI-based Feedback Features used in the interpretable feedback interfaces. Note that we describe them as revised after findings from the formative study (described later). \textbf{Attribution (A) explanations} (Figure \ref{fig6}) are presented as red highlights to indicate key words that ideators should consider changing to improve the ideation scores. Darker reds indicate more negative attributions. These exclude stop words (e.g., ``the", ``to"). To limit information overload, only three words are highlighted. Since the quality and diversity objectives are not necessarily aligned (Figure \ref{fig3}), the highlights are specific to each score only one at a time. Ideators view explanations for each score by selecting its radio button (in table heading). When the ideator hovers her mouse over a highlighted word, a popup will show its attribution sub-score. 

\textbf{Contrastive Attribution (X) explanations} (Figure \ref{fig7}) are similar to Attribution explanations, but only word differences between two iterations are highlighted. The last iteration can be compared against any earlier iteration. Deletions are highlighted in the earlier iteration, while insertions are highlighted in the last iteration. Red highlights have the same meaning as for Attribution explanations, while blue highlights indicate beneficial edits (positive attribution) for verifying successful edits. Darker blues indicate stronger improvements. Currently, only iterations of the same ideation are compared, though future work could compare different ideations. 

\textbf{Counterfactual Suggestion (C) explanations} (Figure \ref{fig8}) suggest alternative words to replace a problematic word (negative attribution). Hovering on a red highlighted word shows a popup with its attribution to the score (same for A and X explanations), and lists related words that could be used for substitution. The potential change in both scores is calculated for each related word. Ideators can decide whether to improve quality or diversity, decide their relevance, and integrate that word or something else. Ideators are reminded that they can propose their own terms instead of what was suggested. To limit confusion and increase utility, only words which a) have an increase in either score and b) are not too irrelevant (i.e., not distant from ``physical activity") are included. 

\subsection{System Implementation and Initialization}
The system was deployed in a survey implemented in Qualtrics, which embeded an external webpage for the ideation user interface, hosted from a web server with Intel Xeon CPU E5-2640 v4 @ 2.40Ghz x 40, 128GB RAM, and Tesla P100 GPU. We used the GPU for calculating the USE embeddings for prompt phrases, ideations, and words; this is used for diversity score predictions and attribution calculations. Each calculation of the score prediction and explanations took 3-4 seconds for messages with 20-40 words, which pilot participants found as an acceptable wait time.

To prompt ideators, we chose 50 Directed phrases from the corpus of \cite{Cox2021}, which we randomly sampled without replacement for each participant. Each ideator will not see repeated prompts. The collection of prior ideations was initialized with 50 ideations from \cite{Cox2021} randomly chosen from the None condition (no prompting). For ecological (external) validity, we dynamically updated the collection of prior ideations after each ideator submits an ideation. This captures the growth of diversity in the collection as more ideations are submitted. We do this separately for each Feedback Interface variant to keep them independent, i.e., ideations from ideators in each UI condition are only added to the collection of prior ideations for that condition so as not to ``contaminate" other collections.

\section{Evaluation}





{\color{black}
We evaluated the usability, understandability, usage and usefulness of different Feedback Interfaces in multiple studies. 
First, we conducted a \ul{Formative Ideation Study} to qualitatively observe how ideators use the different feedback and elicit their opinions about the feedback. This helped us to identify usage patterns, ensure that the interfaces were usable and useful, and mitigate any usability issues.
Next, we conducted a controlled \ul{Summative Ideation Study} with participants as ideators to determine if providing explanations improved ideation Quality and Diversity over No Feedback or Score-only Feedback baselines, to compare which explanations were better, and to assess the ease of understanding and ease of use for each explanation type.
In this study, we measured ideation Quality and Diversity from computational metrics, but these may be non-representative of how people would perceive the ideations. 
Hence, we conducted Validation Studies of \ul{Ideation Quality} and \ul{Ideation Diversity} to measure perceptions from third-party validators. Two studies were required due to the different experiment procedures for each measure --- individual ideation rating for Quality and pairwise ideation rating for Diversity, like in \cite{Cox2021}.
}

\subsection{Formative Ideation User Study}
We conducted a formative user study to investigate 1) how users interpreted various explanation features, 2) how that influenced their ideation, and to identify 3) usability or interpretability issues in our initial design. We then refined our user interfaces for the subsequent summative user study (described later).

\subsubsection{Method and Procedure}
We recruited 15 participants from a university mailing list. They were 6 male, 9 female, with ages 21-33 years old, and all were students. The experiment took 40-50 minutes and participants were compensated \$10 SGD (\$7.43 USD). We employed a within-subjects design with Latin square arrangement to mitigate order effects. We conducted the study online via a Zoom audio call with screen recording, which the participant consented to. Participants completed a tutorial and could clarify instructions with an experimenter. Next, participants performed one ideation session for each of all 6 variants of the Feedback Interface listed in Table \ref{table1}. For each ideation session, the participant was prompted with a different phrase to ideate a motivational message for physical activity. After submitting the attempt, she saw the feedback, and revised her message; this happened for two revisions. Thus, the experiment was within-subjects with 6 Conditions $\times$ 1 Ideation $\times$ 3 Iterations. Using the think aloud protocol, participants were encouraged to speak their thoughts as they read the prompt, feedback, and thought about what to write. We also asked them about their experience with feedback and comments for improvement. 

\subsubsection{Qualitative Findings}
We transcribed the interviews and performed a thematic analysis of usage and utterances. Guided by the three goals for the formative study, one co-author coded the findings into themes with regular discussion with other co-authors. We organized our findings by our three objectives: feedback interpretations, ideation approach, and interpretability issues.

\textit{Feedback interpretations.} Typical feedback usage involved: 1) noting the prediction scores; 2) examining which words were highlighted in red to \textit{``know which words to focus on"}[P13]; 3) looking up related words with the counterfactual suggestion popup, and considering which would lead to the highest score increases for both diversity and quality. After the second iteration, participants would also use the compare mode to check which words were detrimental or useful. In summary, the exploration sequence was \ul{attributions $\rightarrow$ counterfactual suggestions $\rightarrow$ contrastive attributions}. However, participants had some difficulties when making sense of the feedback. Some participants tried to \ul{self-explain} the quality score, since this concept is more commonplace, but had to depend on the feedback regarding the diversity score, since \textit{``without the scores I wouldn't be able to tell whether my sentence is diverse"} [P12] and \textit{``with the related words I know whether there will be an increase"} [P13].

When not receiving explanations, some participants struggled to deeply understand why their ideation scored poorly on quality. For example, P14 found that \textit{``the scores seem a little arbitrary"}; P10 felt that it \textit{``doesn't show why it is motivating"}; and P3 wondered why `exercise' was highlighted
, since \textit{``[exercise] is the main word, not like I could really change anything about it"}. 
This was actually because many participants redundantly used this word. 
Currently, our approach highlights culpable words, but this suggests the \ul{need for semantic explanations}. Participants were also confused when their scores decreased despite following the explanations. P12 \textit{``tried to change the word with a suggestion but wasn’t sure why the score decreased"}. After being suggested the word `desirable' to replace `wanted' with a projected +5\% to quality the score, P13 wrote `desired' and was confused to get a 2\% score decrease instead. Thus, the high dimensionality of language modeling can lead to \ul{spurious errors due to non-linear relations}, harming user experience. Nevertheless, in our later study, we found that the feedback was useful for multiple users. Finally, while our feedback was automated from a corpus and language model, P10 wanted suggestions to \textit{``try including [words] based on other people's answers"} to rely on social proof \cite{Cialdini1999}.

\textit{Ideation approach.} Participants ideated differently based on feedback type. Without explanations, they mostly depended on trial-and-error. P13 felt that writing more specifically, concretely, or with simpler words could increase diversity scores; for example, she revised the term `physical activity' to `pull up' to be more diverse \textit{``because you're taking a specific activity rather than a general term"}. \ul{Writing more specific terms} is consistent with goal setting theory \cite{Locke1990,Locke2002} and distinct words have different embedding placements in our vector space. Participants could be more focused with highlighted words as they \textit{``gave me something to work off"} [P14]. Next, we describe some interesting breakdowns and user mitigations.

Some participants struggled with the potentially divergent nature of the quality and diversity criteria. 
There was diverse preference to focus on either improving quality since it was more intuitive (e.g., P6, P10), or diversity since it fluctuated more at each iteration (e.g., P1, P4).
Hence, to \ul{prioritize either criteria}, each score could be rescaled to nudge users accordingly. 
Due to the breadth of concepts in ConceptNet, participants found that the suggested words sometimes seemed irrelevant, yet some could be \ul{tangentially inspiring}. On being suggested the terms `skate' and `release energy' to replace `exercise', P11 substituted with `swimming', perhaps because of finding another activity that is more energetic than skating and remembering terms starting with `s'. Some participants were \ul{too adherent} to the suggestions to the extent of \ul{losing task relevance}. On receiving the term `arsenic trichloride' (with potential +1\% for Diversity and +4\% for Motivating) to replace her word `organic', P11 used the chemical term in her ideation and rewrote ``... by augmenting physical activities with \MyColorBox[redOrganic]{organic} supplements" to ``... by augmenting physical activities with arsenic trichloride vitamins", which is nonsensical since Arsenic trichloride ($AsCl_3$) is actually a highly toxic substance. This indicates that it is important to have an additional step to filter ideations for safety. Furthermore, while the feedback is helpful to improve scores, we found that some participants drifted away from the prompt phrase. Starting with the phrase ‘right to take care’ and writing ‘You are \MyColorBox[redResponsible]{responsible} for taking \MyColorBox[redCare]{care} of your own health', P8 \ul{fixated} on improving the lower scoring word `responsible', and ended up writing ``You are in \MyColorBox[redControl]{control} of your own health", which inadvertently dropped the word ‘care’; thus, she \ul{neglected her original prompt}, though this did slightly improve her calculated diversity (55\% to 56\%). Finally, we found that our focus on word-based feedback could limit some ideation styles. Interestingly, when ideating without feedback, some participants wrote with a \ul{collective tone}, e.g., ``Let’s go and exercise" [P1], ``We can start our exercise with some stretching" [P4]; but this tone was absent when feedback was provided, and ideations became \ul{neutral and formal}, e.g., "Exercising will reduce chances of you going for surgeries and you can feel better, lose weight and be fitter" [P1].

\textit{Interpretability issues and remedies.} 
{\color{black}We discuss issues identified by participants and remedies that we subsequently implemented for the interfaces tested in the summative study.}
Our attribution explanations originally highlighted about 6 words per ideation and users found it too tedious to track and manage all of them. \textit{Remedy}: we limited the highlights to 3 words with the most negative attributions. Interviewees were confused with many counterfactual words, since they had negative or low improvement scores or were not semantically relevant (e.g., replacing `play' with suggested word `kids' with potential +0\% for Diversity and +0\% for Motivating). \textit{Remedy}: we ensured that suggested words have at least one positive score, and limited to relevant suggestions by eliminating words that had embedding distances too far from words `exercise" and `physical activity". Participants had found the \textit{compare} mode (to show contrastive attributions) useful, but tended to forget to switch over to it. \textit{Remedy}: we set contrastive explanations as default for each iteration if this explanation was available. The default can be gradually reset after users acclimatize to remembering this feature. We implemented these improvements in the feedback UI, and launched a larger summative controlled study to measure the impact of explanations on ideation quality and diversity.

\begin{table*}
\caption{Dependent variables (DV) measured in the summative ideation user study.}
\vspace{-0.2cm}
\label{table2}
\small
\begin{tabularx}{\textwidth}{p{0.15\textwidth}p{0.11\textwidth}X}
\toprule
 Measure 
 & 
 Of & 
 Description and Justification\\
 
 \midrule
 
 Ideation task time & 
 Iteration & 
 Duration to ideate at each iteration. Objectively measures ideation fluency \cite{Barbot2018} or ease of ideation \cite{Cox2021}. 
 \vspace{0.1cm} \\
 
 \grayline
 
 \vspace{-0.15cm} Perceived Importance & 
 \vspace{-0.15cm} Quality score,\newline Diversity score & 
 \vspace{-0.15cm} Ideators’ preference of perceived importance for message quality or diversity 
 {\color{gray}[7-point Likert scale: –3 = ``Motivating much more important", +3 = ``Diverse much more important"]}.\\
 
 \vspace{-0.13cm} Perceived \newline Ideation Quality & 
 \vspace{-0.13cm} Ideation\newline (Final iteration) & 
 \vspace{-0.13cm} Ideators' self-assessment of how motivating and creative the ideation is \newline 
 {\color{gray}[7-point Likert scale: –3 = Strongly Disagree, +3 = Strongly Agree]}.\\
 
 \vspace{-0.13cm} Understanding, Ease of Use, Helpfulness &
 \vspace{-0.13cm} Prompt,\newline Feedback Feature & 
 \vspace{-0.13cm} Ideators' perceptions  asked separately of the Prompt and each Feedback Feature (S of Quality and Diversity, A, X, C) 
 {\color{gray}[5-point Likert scale: –2 = Strongly Disagree, +2 = Strongly Agree]} 
 and rationale {\color{gray}[Open text]}. \\
 
 \vspace{-0.13cm} Usage & 
 \vspace{-0.13cm} Prompt,\newline Feedback & 
 \vspace{-0.13cm} Qualitative description of how the feedback information was used {\color{gray}[Open text]}.\newline This was only asked of the 2nd ideation for each interface section.
 \vspace{0.1cm} \\
 
 \grayline
 
 \vspace{-0.13cm} Computational\newline Diversity & 
 \vspace{-0.13cm} Ideation & 
 \vspace{-0.13cm} Metrics of diversity based on ideation embeddings \cite{Cox2021} 
 {\color{gray}[Ideation Dispersion (MST Mean of Edge Weights), Ideation Disparity (Mean Pairwise Distance), and Repeller Chamfer Distance (Mean Min Pairwise Distance)]}. \\
 
 \bottomrule
\end{tabularx}
\end{table*}

\subsection{Summative Ideation User Study}
We conducted a mixed-design experiment with main independent variable (IV), Feedback Condition (N, S, SA, SAX, SAC, SAXC) and secondary IV, Iteration (t=1,2,3). Iteration was fully within-subjects, while Feedback Condition was partially within-subjects. For each ideation, participants saw a Directed prompt from a pre-selected set of 50 (Section 4.3) \cite{Cox2021}. All Feedback Conditions had the same 50 prompts. We exposed each participant to 2 randomly-selected Feedback Conditions with 2 prompts per type and 3 iterations per prompt (total 12 ideation iterations) to mitigate individual variance, while limiting fatigue with too many trials. We limited the prompt to contain only one phrase, since users could struggle to utilize multiple phrases \cite{Cox2021}. While the system could show explanations for both Quality and Diversity, we showed Diversity explanations by default to prioritize improving diversity which users may find less intuitive (Section 5.1.2). Table \ref{table2} describes dependent variables measured. The experiment apparatus and survey questions were implemented in Qualtrics (see Appendix A.2.1).

\subsubsection{Experiment Task and Procedure}
Participants were tasked to write motivational messages towards physical activity using various feedback with the following procedure:
\begin{enumerate}[leftmargin=.5cm]
    \item[1.] \ul{Introduction} to experiment objective and consent to the study.
    \item[2.] \ul{Screening quiz} with a 4-item word association test \cite{Chandler2019} to assess English language skills. 
    All answers must be correct to continue.
    \item[3.] \ul{UI section} ($\times 2$) with different Feedback Conditions.
    \begin{enumerate}
        \item[a)] \ul{Tutorial} about ideation process, and how to use the feedback.
        \item[b)] \ul{Ideation session} ($\times 2$) to ideate based on a prompt each.
        \begin{enumerate}
            \item[1.] \ul{Prompted ideation} to view a prompt, write an initial ideation in one to three sentences, and submit for automatic review. This page is timed to measure ideation task time.
            \item[2.] \ul{Iterated revision} ($\times 2$) to receive feedback and revise.
            \begin{enumerate}
                \item[1.] \ul{Prompted ideation} to view the same prompt again.
                \item[2.] \ul{Ideation feedback} to inform the participant where and how to improve their previous ideation.
                \item[3.] \ul{Ideation revision} and submission. With these two revisions, there are three ideation iterations.
            \end{enumerate}
            \item[3.] \ul{Perception questionnaire} to ask participants about their perceptions of usage and usability (Table \ref{table2}).
        \end{enumerate}
        \item[4.] \ul{Post-questionnaire} on demographics.
    \end{enumerate}
\end{enumerate}
To control fatigue, we limited to 4 ideation trials. Instead of 4 ideations $\times$ 1 condition (no repeated trials, potentially noisy), or 1 ideation $\times$ 4 conditions (fully between-subjects, high individual variance), we balanced with 2 ideations $\times$ 2 conditions.

\subsubsection{Experiment Data Collection}
We recruited participants from Amazon Mechanical Turk (AMT) with high qualification ($\geq$5000 completed HITs, >97\% approval). 104 workers attempted the survey, 97 passed the screening quiz, and 70 completed the survey. They were 42.0\% female, between 23 and 66 years old (M=39.5). Participants were compensated US\$4.00 after completing the ideation tasks and surveys. Participants completed the survey in median time 26.0 minutes and were compensated ~US\$9.24/hour. We collected 4 ideations per participant (2 Feedback Conditions x 2 Ideations) with up to 3 Iterations each, from 70 participants for a total of 280 ideations across the 6 Feedback Conditions. 

\subsubsection{Statistical Analysis}
We fit linear mixed effects regression (LMER) models on various dependent variables, performed ANOVAs on the fixed main and interaction effects, and post-hoc contrast tests for specific differences identified. {\color{black}Since LMER models can accommodate missing data by imputing and estimating values, we can analyze our experiment with partially within-subjects independent variables.} Due to the large number of comparisons in our analysis, we consider differences with p<.001 as significant and p<.005 as marginally significant. This is stricter than a Bonferroni correction for 50 comparisons (significance level = .05/50).

For \textbf{perceived ratings} (Usefulness, Ease of Use, Understandability), we fit a LMER model with Feedback Feature (Prompt, Diversity Score, Quality Score, Attribution, Contrastive, Counterfactual) as fixed effect, and Participant as random effect (Table \ref{table5} in Appendix A.3). Note that the features were considered separately rather than as a combination in the Feedback Interface.

\begin{figure*}
  \centering
  \includegraphics[width=0.81\linewidth]{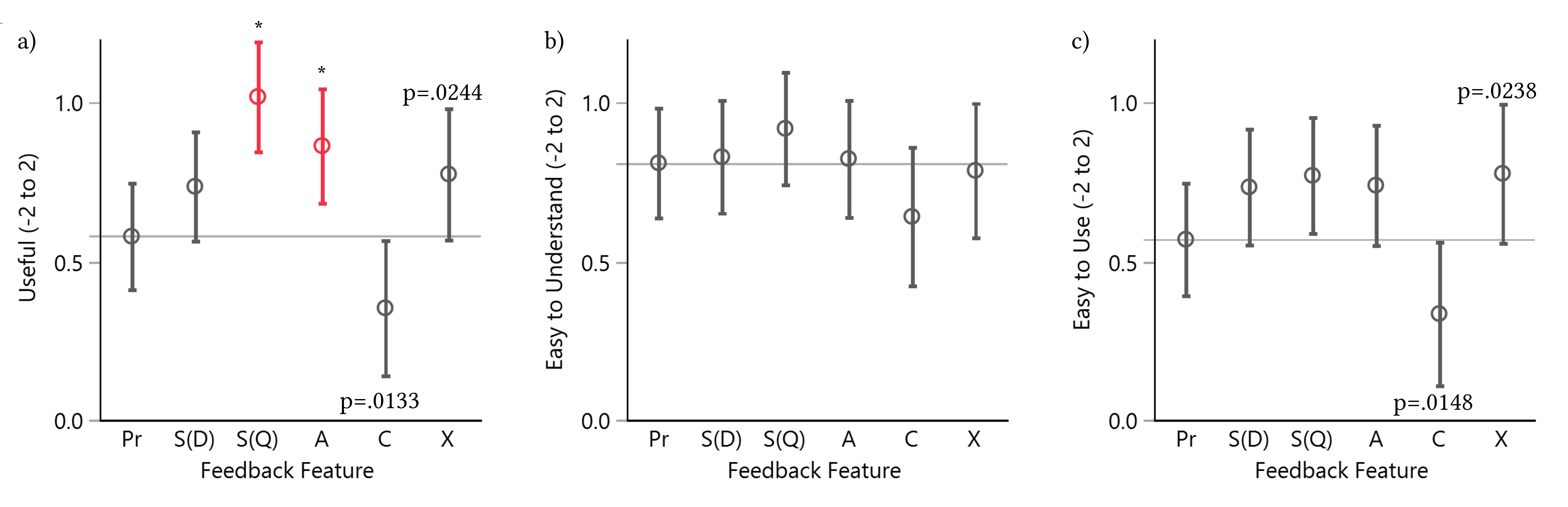}
  \vspace{-0.48cm}
  \caption{Results of ideators’ perceived ratings for each feedback feature. Feedback Features: Pr = Prompt, S(D) = Diversity Score, S(Q) = Quality Score, A = Attribution explanation, C = Counterfactual suggestion, X = Contrastive attribution. Red markers indicate significant differences from Pr. p values indicate contrast tests relative to Pr. * indicates p<.0001.}
  \Description{Three sub-figures of interval plot of user ratings for each feedback feature regarding usefulness, ease to understand, and ease to use.}
  \label{fig9}
\end{figure*}

\begin{figure*}
  \vspace{-0.2cm}
  \centering
  \includegraphics[width=0.85\linewidth]{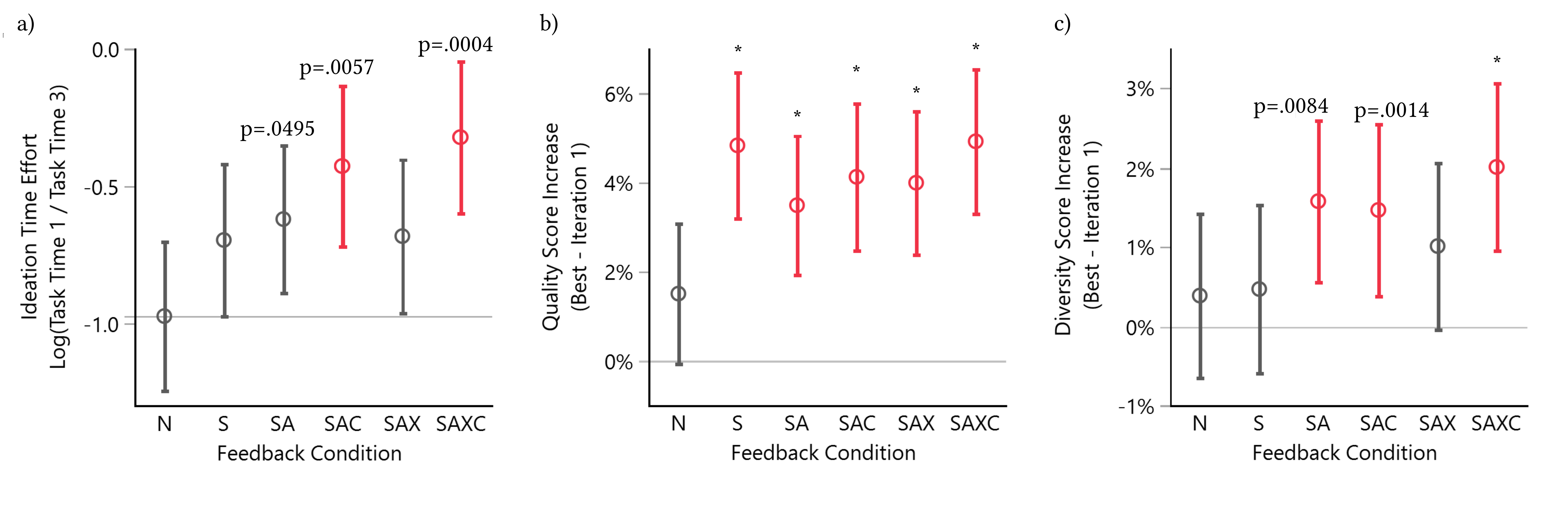}
  \vspace{-0.65cm}
  \caption{Results of ideator performance. a) Time effort ratio between Iteration 1 and 3. b) Increase in Quality Score from Iteration 1 to the iteration with best Quality Score. c) Increase in Diversity Score from Iteration 1 to the iteration with best Diversity Score.
  Feedback Conditions: N = None, S = Score Prediction, SA = Score + Attribution, SAC = SA + Counterfactual, SAX = SA + Contrastive, SAXC = SA + X + C. p values indicate 1-sample, 1-tail t-tests relative to N (a) or 0\% (b, c); 
  * indicates p<.0001.
  }
  \Description{Three sub-figures of interval plot for each feedback condition regarding ideation time, the increase of ideation quality score, and the increase of ideation diversity score.}
  \label{fig10}
\end{figure*}

\begin{figure*}
  \vspace{-0.2cm}
  \centering
  \includegraphics[width=0.85\linewidth]{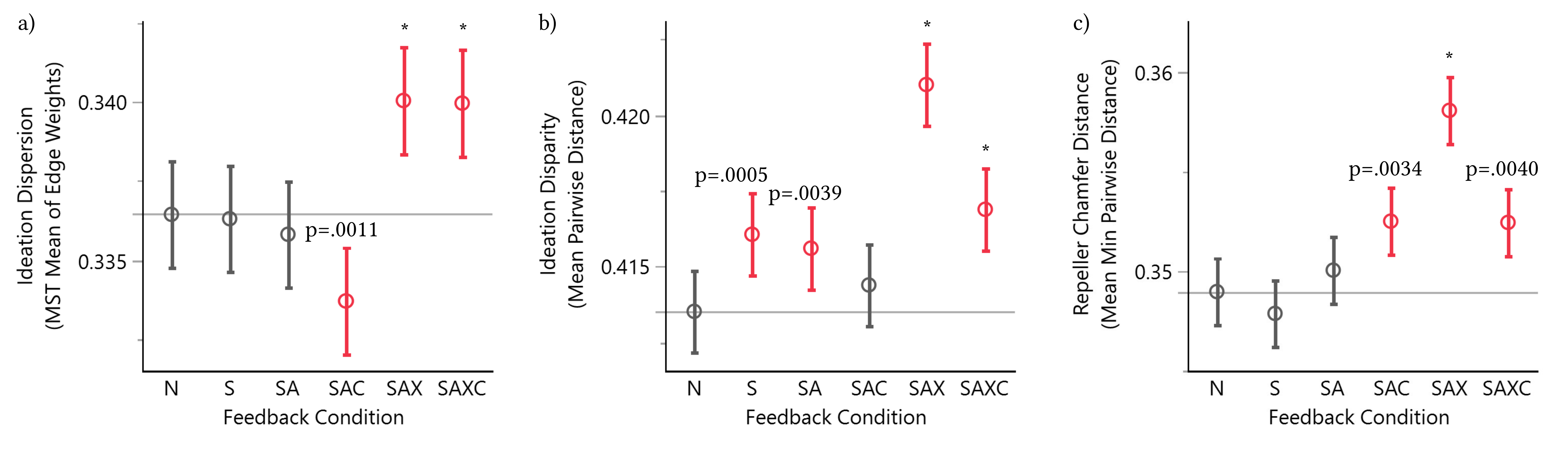}
  \vspace{-0.4cm}
  \caption{Ideation Diversity calculated with different computational metrics: a,b) diversity of new and prior ideations, and c) new ideations from prior ones. p values indicate contrast tests relative to N. * indicates significantly different from N at p<.0001.}
  \Description{Three sub-figures of interval plot for each feedback condition regarding three computational metrics of Ideation Diversity, including Ideation Dispersion, Ideation Disparity, and Repeller Chamfer Distance.}
  \label{fig11}
\end{figure*}

For \textbf{ideation performance}, we examined ideation speed, and whether Diversity and Quality Scores increased. We fit a LMER model with Feedback Condition and Prompt (which one of the 50 prompts was shown) as fixed main effects, and Participant as random effect (Table \ref{table6} in Appendix A.3). We evaluated whether ideation took longer with Score and XAI feedback compared to the baseline 1st iteration,\enlargethispage{10pt} so we calculated the ratio between the 1st and 3rd iteration with a log transform (to improve normality) to get the Ideation Time Effort metric. Note that the best Score may be achieved at the 2nd or 3rd iteration, i.e., the participant may inadvertently get a lower score on the 3rd attempt. Hence, we calculated the difference between the best Score of the latter two iterations with the first, and did so separately for Diversity and Quality. There was a strong negative correlation (r = –.436, p<.0001) between the Diversity and Quality Scores, so we included Quality Score Increase as a fixed main effect too for the Best Diversity Score Increase response, and vice versa. We also tested the Score Increase estimates from the LMER model against no increase with a 1-sample, 1-tail t-test for each Feedback Condition against the constant 0.\enlargethispage{10pt}

To analyze objective \textbf{diversity} for each Feedback Condition, we resampled 50 bootstrap samples, calculated three computational diversity metrics defined in \cite{Cox2021} (Ideation Dispersion, Ideation Disparity, and Repeller Chamfer Distance), and fit a LMER model with Feedback Condition as fixed main effect. In Appendix A.3, Table \ref{table5}, Table \ref{table6}, and Table \ref{table7} summarize the goodness of fit (R2) of the LMER models, and statistical significance of ANOVA tests.\enlargethispage{10pt}

\subsubsection{Results}
We describe participant priorities and perceptions of their ideations and the feedback, and their performance as calculated with metrics. When ideating, participants prioritized improving their Quality Score (56.5\%, rating < 0) instead of Diversity Score (12.6\%, rating > 0), and 31\% balanced between the two. They perceived their ideations as Motivating (high quality, 81.3\% with rating > 0) and Creative (73.8\%), though there was no significant difference across Feedback Conditions. Participants rated all features as useful, easy to understand and use, except for Counterfactual suggestions which had lower usefulness and ease of use (Figure \ref{fig9}).

As expected, participants took more time to ideate when receiving Score (S) and XAI feedback, and significantly so for SAC and SAXC compared to None (Figure \ref{fig10}a); suggesting that they were studying these feedback. S feedback increases were non-significant or marginal compared to no feedback (None). However, compared to no improvement (Increase = 0), we found that showing S or all explanations significantly improved Quality Score (Figure \ref{fig10}b), and showing SA, SAC, and SAXC significantly improved Diversity Score (Figure \ref{fig10}c). S did not improve Diversity Score, perhaps since ideators were blind to what they could change or do without explanations, and would end up randomly editing. Explanations did not improve Quality compared to just showing Scores (p=n.s.), and only very marginally improved Diversity (p=.0414 for SAXC). The lack of differences could be due to the high variability in ideations and limited number of participants; future work could recruit more than 50 participants per condition.

The three computational diversity metrics \cite{Cox2021} — Ideation Dispersion, Ideation Disparity, and Repeller Chamfer Distance —  measure slightly different aspects of diversity, and had different trends (Figure \ref{fig11}). The first two metrics calculate the total diversity of combining new ideations with the seed prior ideations, and the third metric measures how different the new ideations are from the prior ones. In general, SAX had the highest diversity. Other results were somewhat indeterminate. SAC had the lowest Ideation Dispersion, but higher Repeller Chamfer Distance; this suggests that Counterfactuals may help diversify ideations from those without any feedback (N), but may not diversify ideations from other forms of feedback. The detrimental effect of Counterfactual suggestions may also hinder diversity when combined with Contrastive explanations (SAXC; Figure \ref{fig11}b,c).

We qualitatively analyzed ideator rationales to interpret their perceived usefulness, understanding, and usage of various ideation feedback. Many findings echo our earlier ones from the formative study. Participants were mixed regarding whether Score feedback was helpful. P36 \textit{``found the feedback easy to understand because it gave me a number to see how well my [message] is."}; but P13 \textit{``found the feedback difficult to understand because I have no idea how it's calculated. I can't tell what part of my idea it's targeting so I have no idea how to improve. It feels random."}; P46 was more positive, finding them \textit{``somewhat helpful, but did not give guidance on how to improve my scores"}. This validates the \ul{need for deeper explanations}.

However, the receptivity towards Attribution feedback was also mixed. P41 reported that \textit{``feedback was good, preventing me from overusing common phrases"}, and P43 \textit{``did like the highlights to look for different wording that is not so generic".} The highlights were more useful when used in contrast mode to show Contrastive attributions, e.g. \textit{``The compare mode definitely showed me which words I should keep and which ones I should edit, to get the optimal score."} [P68]. In contrast, P64 wanted more useful explanations and found that attribution \textit{``colors and words are not as helpful as they really are ambivalent."} P63 felt \textit{``it was often not clear why a certain word would be less valuable than another or how it would affect the scores, especially diversity."} These cases indicate highlighting salient words and rating their attribution \ul{lacks rationalization insights} \cite{Ehsan2018}.

Perceptions towards Counterfactual suggestions were more negative, though sometimes positive. P22 found ideating \textit{``easier now that there were suggested words", and P07 said it ``helped me to add words to increase inspiration."} However, others found the suggested words problematic: P70 felt that \textit{``sometimes the suggested words seemed completely unrelated to the prompt or even fitness itself."}; P28 felt that \textit{``the suggestions make the phrase a lot worse, non-grammatically correct"}; P32 \textit{``found the feedback a little confusing because it seemed like when I changed words to try to increase my score it ended up keeping it close to the [previous] score."} Hence, \ul{Counterfactual suggestions should be relevant, grammatical, and efficacious}.\enlargethispage{10pt}

Providing all explanations together enabled rich usage. P22 \textit{``changed words based on suggested words, excised words that complicated the message needlessly, and used the comparison checker to see if there was a word that particularly hurt between each prompt."} Therefore, it is important to provide these explanations together.

\begin{table*}
\caption{Dependent variables (DV) measured in the ideation quality and diversity validation user studies.}
  \label{table3}
  \vspace{-0.1cm}
  \small
\begin{tabularx}{\textwidth}{p{.195\textwidth}p{.085\textwidth}X}
\toprule
 Measure & Of & Description and Justification \\
 \midrule
 Individual Quality\newline (Motivating/Informative/Helpful)
 & Ideation & Rating of how the message feels {\color{gray}[7-point Likert scale: –3 = Very Demotivating/Uninformative/Unhelpful, +3 = Very Motivating/Informative/Helpful]}. \\
 \grayline
 Pairwise Dissimilarity& Ideation-Pair & Rating of perceived difference between 2 ideations {\color{gray}[20-interval slider: 0 = Identical, 100 = Highly different]}. \\
\bottomrule
\end{tabularx}
\end{table*}

\begin{figure*}
  \centering
  \vspace{0.4cm}
  \includegraphics[width=\linewidth]{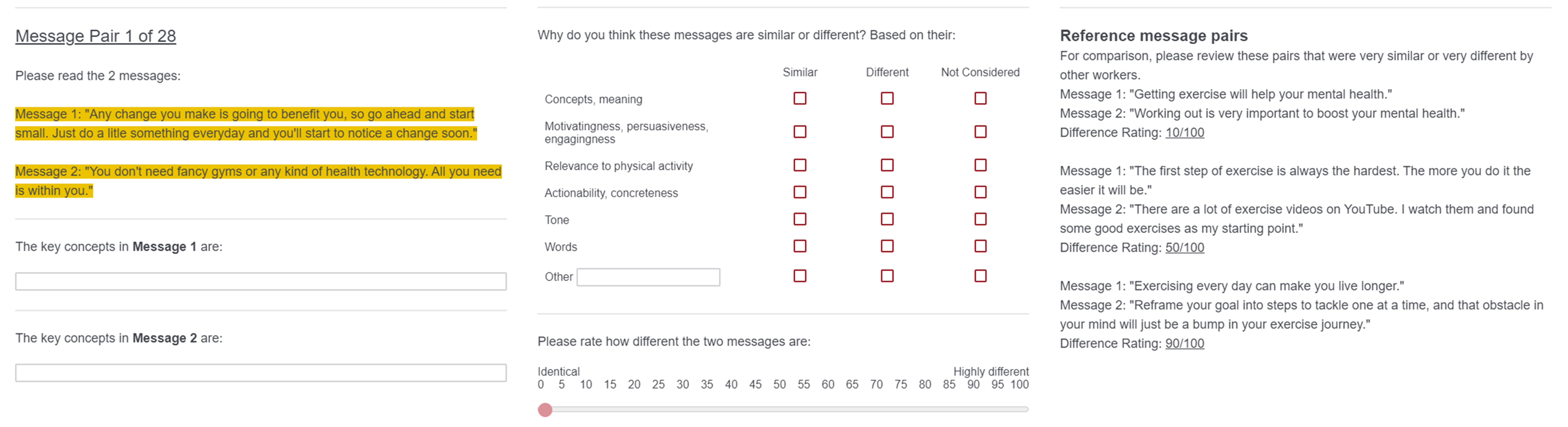}
  \vspace{-0.5cm}
  \caption{Example trial in Ideation Diversity Validation User Study for participants to examine a pair of ideations (Left), rationalize and rate their dissimilarity (Middle), while referring to three reference pairs with varying dissimilarities.}
  \Description{Screenshots of the survey in Ideation Diversity Validation User Study.}
  \label{fig12}
\end{figure*}

\subsection{Ideation Quality Validation User Study}
To evaluate ideation quality, we conducted a within-subjects experiment with Feedback Condition as independent variable. Similar to \cite{Cox2021,kocielnik_send_2017}, we measured quality with three 7-point Likert scale questions on whether an individual ideation was Motivating, Informative, or Helpful (Table \ref{table3}). The ideations evaluated included the third iteration of all ideations for each Feedback Conditions from the Summative Ideation User Study (N=43 to 49; 278 total).

\subsubsection{Experiment Task and Procedure}
Participants were tasked to \textit{read} motivational messages towards physical activity, and rate their quality (Appendix A.2.2). Participants went through the same procedure as in the Ideation User Study, but with a different Step 3:
\begin{enumerate}[leftmargin=.5cm]
    \item[3.] Assess 25 messages regarding how well they motivate for physical activity. For each message,
    \begin{enumerate}
        \item[a.] Read a random message from one of the 6 Feedback Conditions (N, S, SA, SAC, SAX, SAXC).
        \item[b.] Rate on a 7-point Likert scale, whether the message is motivating (effective), informative, and helpful.
        \item[c.] Write their rationale for their quality ratings. This was only asked randomly two times to mitigate response fatigue with texts that are too short or copied from previous responses.
    \end{enumerate}
\end{enumerate}

\subsubsection{Experiment Data Collection}
We recruited participants from AMT with the same high qualification as the ideation study. 71 workers attempted the survey, 60 passed the screening task and complete the survey (84.5\% pass rate). They were 41.7\% female, between 25 and 61 years old (M=36.0). Participants completed the survey in median time 12.1 minutes and were compensated US\$1.50. In total, 278 messages were rated 1500 times (M=5.40x/message); 2 omitted by random chance. Average aggregate-judge correlations \cite{chan2017semantically} for motivation, informativeness, helpfulness ratings were r=.472, .467, .469, respectively, indicating reasonable inter-rater agreement.

\subsubsection{Statistical Analysis and Results}
We binarized ratings to evaluate as continuous variables and fit LMER models on three quality ratings like in the Ideation User Study with Feedback Condition as fixed effect, and Participant and Ideation as random effects (Table \ref{table8} in Appendix A.3). Ideations were rated as Motivating (M = 72.6\%), Informative (M = 70.1\%), and Helpful (M = 75.5\%), but there were no significant differences across Feedback Conditions (all p=n.s.).

\subsection{Ideation Diversity Validation User Study}
The ideation study measured diversity with computational metrics, but people may be less sensitive to perceiving diversity, so it is necessary to evaluate with a user study. Here, we ask a new set of third-party participants to validate ideation diversity from each Feedback Condition. We conducted a mixed-design experiment with Feedback Condition (4 levels: N, S, SAX, SAXC) as independent variable. We omitted SA and SAC, which did not improve computational diversity (Figure \ref{fig11}) and were thus unlikely to increase perceived diversity. Similar to \cite{Cox2021,Dow2010,Siangliulue2015}, participants rated the perceived dissimilarity between a pair of ideations (Table \ref{table3}). From the ideation study, one collection of up to 100 ideations was obtained for each Condition consisting of 50 prior messages (from \cite{Cox2021}) and up to 50 new ideations (3rd iteration) from participants. Similar to \cite{Cox2021}, we randomly sampled 200 ideation pairs per condition; each pair could contain 2 prior, 2 new, or 1+1 mixed ideations.

\subsubsection{Experiment Task and Procedure}
Participants were tasked to review pairs of motivational messages, and rate their dissimilarity (Appendix A.2.3). The experiment procedure is as follows:
\begin{enumerate}[leftmargin=.5cm]
    \item[1.] \ul{Introduction} to experiment objective and \ul{consent} to the study.
    \item[2.] \ul{Screening quiz} which needs all correct answers with
        \begin{enumerate}
            \item[a)] 4-item \ul{word association test} \cite{Chandler2019} of English language skill,
            \item[b)] \ul{Similarity judgment test} on identifying the most similar ideation pair from 3 pairs (set to be easy). 
        \end{enumerate}
    \item[3.] \ul{Practice session} with three pair trials. For each trial:
        \begin{enumerate}
            \item[a)] \ul{Read} two ideation \ul{messages}.
            \item[b)] \ul{Write} key \ul{concepts} for each message as attention check.
            \item[c)] \ul{Rationalize} the pair dissimilarity by multiple categories (identified from a pilot study with 660 rationales from 110 participants). This is to prime rigorous study of the messages and mitigate mindless rushing. We designed the question as a matrix of checkboxes (Figure \ref{fig12}, Middle Top) instead of free text, to reduce response fatigue.
            \item[d)] \ul{Rate perceived dissimilarity} on a 20-interval slider from 0 to 100 (Figure \ref{fig12}, Middle Bottom). We used a high-resolution slider instead of a Likert scale for more precise measurement.
            \item[e)] \ul{View the ``correct" dissimilarity rating} of 3 practice trials with ratings 1, 5, 9.
            The rating values were averages from pilot study participants, which we use to calibrate participants towards more consistent ratings.
        \end{enumerate}
    \item[4.] \ul{Main session} with 28 message pairs (7 pairs per Feedback Condition, 4 Feedback Conditions). Same procedure as the Practice session, except for omitting Step 3e, and for Step 3d:
        \begin{enumerate}
            \item[d)] \ul{Rate perceived dissimilarity} with the same interface as Step 3d and reference of the 3 practice pairs with "correct" ratings to anchor participants (Figure \ref{fig12}, Right).
        \end{enumerate}
    \item[5.] \ul{Post-questionnaire} on demographics.
\end{enumerate}

\subsubsection{Experiment Data Collection}
We recruited participants from AMT with the same high qualification as the ideation study.  162 workers attempted the survey and 125 passed screening and completed the survey (77.2\% pass rate). They were 41.6\% female, between 20 and 67 years old (M=34.0). Participants completed the survey in median time 47.6 minutes and were compensated US\$6.00. In total, 800 message pairs were rated 3,500 times (M=4.38x / pair). The average aggregate-judge correlation \cite{chan2017semantically} was r=.478 for perceived dissimilarity, indicating reasonable inter-rater agreement.

\begin{figure}[t]
  \centering
  \includegraphics[width=0.65\linewidth]{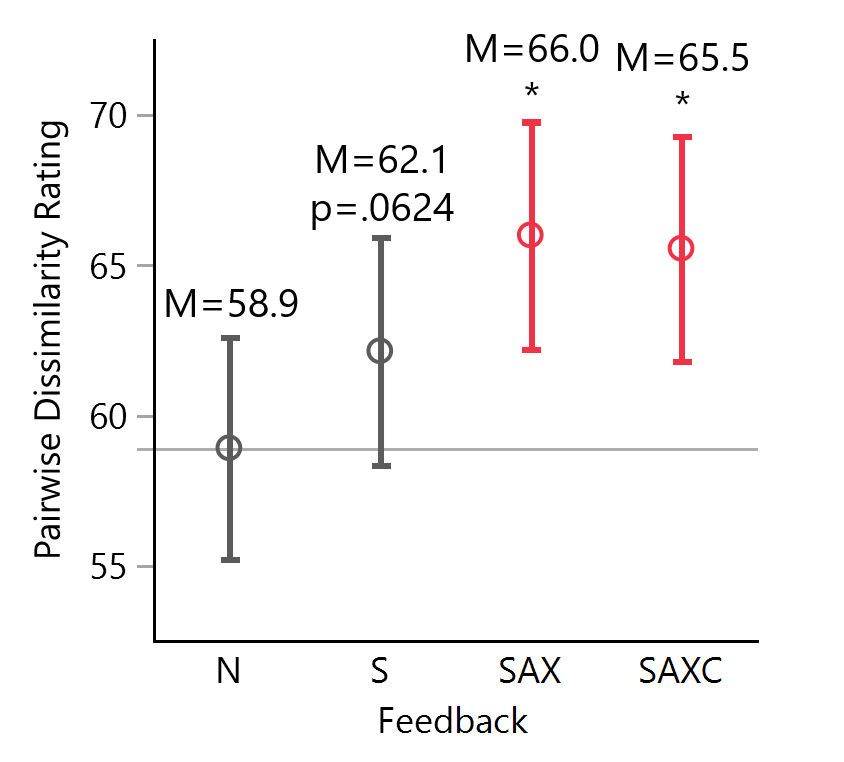}
  \vspace{-0.4cm}
  \caption{Results of validator perceived dissimilarity between ideation pairs from different Feedback Conditions. 
  p values indicate difference from N. * indicates p<.0001}
  \Description{Interval plot of the validator perceived dissimilarity between ideation pairs from different Feedback Conditions.}
  \label{fig13}
  \vspace{-0.3cm}
\end{figure}

\begin{table*}[!b]
  \caption{Summary of results of explanation effects. Grey 0 indicates baseline or not significantly different from baseline (N or Pr.), Grey arrow indicates marginally significant from N, black arrow indicates significant at p<.0001.}
  \vspace{-0.2cm}
\label{tableSummary}
\small
\begin{tabularx}{.915\textwidth}{l l l l cccccc}
 \cmidrule{1-10}
 &  &  &  & \multicolumn{6}{c}{Feedback Interface} \\
 \cmidrule{5-10}
 Effect & Measure & User Study & Evidence & N  & S & SA & SAC & SAX & SAXC \\
 \cmidrule{1-10}
 Ideation Quality & Quality Score Increase & Summative Ideation & Fig. \ref{fig10}b & \textcolor{verylightgray}{0} & $\uparrow$  & $\uparrow$  & $\uparrow$  &  $\uparrow$ & $\uparrow$ \\
  & Motivating/Informative/Helpful ratings & Quality Validation & - &  \textcolor{verylightgray}{0} & \textcolor{verylightgray}{0}  &  \textcolor{verylightgray}{0} &  \textcolor{verylightgray}{0} &  \textcolor{verylightgray}{0} & \textcolor{verylightgray}{0} \\
\grayline
 Ideation Diversity & Diversity Score Increase  & Summative Ideation & Fig. \ref{fig10}c & \textcolor{verylightgray}{0}  & \textcolor{verylightgray}{0}  & \textcolor{verylightgray}{$\uparrow$}  & \textcolor{lightgray}{$\uparrow$}  & \textcolor{verylightgray}{0}  & $\uparrow$ \\
 & Computational Ideation Dispersion &  & Fig. \ref{fig11}a  & \textcolor{verylightgray}{0}  & \textcolor{verylightgray}{0}  & \textcolor{verylightgray}{0}  & \textcolor{lightgray}{$\downarrow$} &  $\uparrow$ & $\uparrow$ \\
   & Computational Ideation Disparity &  & Fig. \ref{fig11}b & \textcolor{verylightgray}{0} &  \textcolor{lightgray}{$\uparrow$} &   \textcolor{lightgray}{$\uparrow$}  &  \textcolor{verylightgray}{0}  & $\uparrow$   & $\uparrow$  \\
   & Computational Repeller Chamfer Distance &  & Fig. \ref{fig11}c & \textcolor{verylightgray}{0}  & \textcolor{verylightgray}{0}  & \textcolor{verylightgray}{0}  & \textcolor{lightgray}{$\uparrow$}  & $\uparrow$  & \textcolor{lightgray}{$\uparrow$} \\
   & Validator pairwise dissimilarity rating & Diversity Validation &  Fig. \ref{fig13} &  \textcolor{verylightgray}{0}  &  \textcolor{verylightgray}{0}  &  \textcolor{verylightgray}{-}  & $\uparrow$  &  \textcolor{verylightgray}{-}  & $\uparrow$ \\
\grayline
 Ideation Effort & Log(Task Time 1 / Task Time 3) & Summative Ideation & Fig. \ref{fig10}a & \textcolor{verylightgray}{0}  & \textcolor{verylightgray}{0}  & \textcolor{verylightgray}{0}  & \textcolor{lightgray}{ $\uparrow$}  &  \textcolor{verylightgray}{0} &  $\uparrow$ \\
\vspace{0.0cm}\\
 \cmidrule{1-10}
 &  &  &  & \multicolumn{6}{c}{Feedback Feature} \\
 \cmidrule{5-10}
 Effect & Measure & User Study & Evidence & Pr.  & S(D) & S(Q) & A & C & X \\
 \cmidrule{1-10}
Ideation Effort & Usefulness rating  & Summative Ideation & Fig. \ref{fig9}a & \textcolor{verylightgray}{0} &  \textcolor{verylightgray}{0}  & $\uparrow$   & $\uparrow$   & \textcolor{verylightgray}{0}   & \textcolor{verylightgray}{0}  \\
  & Ease of Understanding rating &  & Fig. \ref{fig9}b & \textcolor{verylightgray}{0}  &  \textcolor{verylightgray}{0}  &  \textcolor{verylightgray}{0}  & \textcolor{verylightgray}{0}   &  \textcolor{verylightgray}{0}  & \textcolor{verylightgray}{0}  \\
        & Ease of Use rating &  & Fig. \ref{fig9}c & \textcolor{verylightgray}{0}  & \textcolor{verylightgray}{0}   & \textcolor{verylightgray}{0}   &  \textcolor{verylightgray}{0}  &  \textcolor{verylightgray}{0}  & \textcolor{verylightgray}{0}\\
 \cmidrule{1-10}
\end{tabularx}
\end{table*}

\subsubsection{Statistical Analysis and Results}
Similar to our analysis in the summative ideation study, we fit a LMER model on perceived dissimilarity with Feedback Condition, Trial Index (in survey), \# Rationales Considered (indicating thoughtfulness) as fixed effects, and Participant as random effect (Appendix A.3 Table \ref{table9}). Since all Conditions contain pairs with only prior messages, we excluded these pairs since they will have identical dissimilarity across Conditions. 

Ideation message pairs from the SAX and SAXC conditions had higher perceived pairwise dissimilarity than pairs from N (Figure \ref{fig13}).  Pairs from the S condition only had marginally higher ratings than N. This mostly agrees with our results of the computational diversity metrics (Figure \ref{fig11}). Therefore, Attribution and Contrastive Attribution improved diversity, while Counterfactual Suggestion had limited effect.

\subsection{Summary of Evaluation Results}
We summarize our findings from the multiple experiments here and in Table \ref{tableSummary}.
{\color{black}
1) From the formative ideation study, participants wanted some feedback on their ideations, appreciated all XAI types in feedback more than just showing Scores, had a tendency to prioritize either for Quality or Diversity, might have fixated on prompts or suggestions too much, and sometimes struggled with problematic Counterfactual Suggestions.
2) From the summative ideation study, all feedback types were rated equally understandable and useful, though Score and Attribution feedback were rated more useful than showing Prompts only.
3) From the summative ideation and both validation studies, a) Quality Score increased with all feedback types, but this was not perceptible with third-party validators; b) Diversity Score, computed diversity, and validator perceived pairwise dissimilarity were increased for feedback with combined explanations (SAX, SAXC). Therefore, Score, Attribute and Contrastive explanations were especially useful. Counterfactual Suggestions were somewhat detrimental, but this was not significant.
}

\section{Discussion}
We discuss the value of ideation XAI feedback, future improvements, generalization, and implications to human-AI collaboration.

\subsection{Effectiveness of XAI Feedback for Ideation}
We have shown that XAI provides helpful feedback to improve collective creativity in ideation, and multiple explanations should be combined to amplify insights for ideators. Towards improving diversity, SAX feedback was most effective, followed by SAXC which was similarly effective. Counterfactual suggestions were promising but harmful; we discuss improvements later. However, Score feedback alone was very marginally useful and explanations were less effective at improving quality, and we discuss improvements later.

Despite providing helpful explanations, some ideators still struggled to apply feedback for ideation. 
We had suggested the need for semantic explanations rather than correlated attributions (Section 5.1.2). To further help, we could leverage strategies in education: providing examples and self-explanation. We could improve the tutorials and provide worked examples \cite{Atkinson2000} of past ideators deriving new ideations from each explanation feature, particularly including their rationale. Ideators could reflect more rigorously on the feedback by self-explaining \cite{Vanlehn2016} what they learned from the feedback and how they could or could not apply it to edit their ideation.

\subsubsection{Improving XAI Feedback for Ideation Quality}
We had mixed results regarding the impact of Score and XAI feedback on Quality. Feedback did increase the Quality Score in the ideation study, but not the perceived quality in the validation study, perhaps due to several reasons. 
1) Ideators were confounded with a conflicting goal to increase Diversity. Diversity and Quality Scores were highly negatively correlated, and including either as a factor in the LMER models led to significant effects. Nevertheless, we focused on increasing Diversity while maintaining Quality, rather than mainly increasing Quality. For applications prioritizing Quality, ideators could be told to only improve Quality, but at the cost of more redundancy. 
2) Ideators may have felt self-sufficient and not have relied much on XAI to improve Quality. 
3) The Quality prediction model may need to be more accurate to represent and explain Quality, perhaps with a regression model instead of binary classifier. 
4) Although we had used the Universal Sentence Encoder (USE) \cite{Cer2018} that models whole sentences rather than just words, explanations were framed with individual words. This caused ideators to fixate on each word separately rather than longer phrases or gestalt concepts. Presenting the explanation as a knowledge graph, highlighting multiple associated words, or inferring and presenting associated concepts \cite{Zhang2022} may mitigate this fixation. Also, instead of using generic language models, the modeling could be domain-specific to motivation and physical activity and include knowledge bases (e.g., \cite{Bosselut2020,Petroni2020}).

\subsubsection{Providing More Useful Counterfactual Suggestions}
Counterfactual explanations need to be more usable by improving the relevance and grammatical context of suggested words. The grammar was due to lemmatizing attribution words to their root and finding related words in ConceptNet. The suggested words should be converted to fit the grammatical form of the word being replaced. 

The irrelevance of suggested words could be due to ConceptNet using domain-independent data sources (e.g., Wikipedia, Wiktionary, WordNet) \cite{Speer2017}. Words are thus too distant from the domain of physical activity. At the potential cost of stifling creativity by reducing too many suggestions, we had limited the embedding distances of the words (remedy in Section 5.1.2), yet the remaining words were still insufficiently relevant. Future work should balance irrelevance and inspiration by carefully controlling the distance threshold and training ideators to be more comfortable with less relevant words. We can also leverage recent advances in synthetic data generation and use generative adversarial networks (GAN) to synthesize counterfactual sentences that increase quality and diversity (e.g., \cite{Xu2018}). However, these sentences are likely to be ungrammatical or strange, needing crowdworkers to clean up.

Some participants had deviated from the prompt phrase, reducing the precision of the directing. To mitigate this, we can limit the distances of suggested words from the prompts. Reminders can also be sent if the new ideations are measured to be too distant from the prompts by embedding distance. Nevertheless, some deviation can accommodate inaccuracies in the embedding distance calculations.

\subsection{Generalization to Other Applications, Domains, and Explanations}
By bridging the gap between the education-based feedback \cite{Hattie2007} and philosophy-driven AI explanations \cite{Miller2019}, we proposed and evaluated an explainable AI-based ideation feedback system to improve quality and diversity in iterative crowd ideation. Our evaluation on one use case of motivational messaging for physical activity may limit the generalization of our results. We had focused on a low-skill task of everyday writing where lay users typically do not need sophisticated reasoning. In domains requiring expertise, ideators may further appreciate complex explanations and may even better leverage simpler explanations (e.g., plain attribution). For tasks that are non-text-based, such as poster design \cite{Dow2010}, attribution may be ambiguous and counterfactuals more complex to develop or describe. Therefore, explanations will need to be more carefully designed for more complex ideation tasks.

Our approach can be applied to other applications, such as story writing and graphic design. The short length of motivational messages makes computing on and reviewing them tractable, but processing longer documents like short stories \cite{Clark2018} will require larger language models \cite{Devlin2018}. The long documents could also be divided to be iterated one part at a time \cite{bernstein_soylent:_2010}, or summarized automatically \cite{Gambhir2017} before predicting a score and providing simplified feedback. For graphical ideation tasks like graphic design \cite{Oppenlaender2020a} or mood boards \cite{Koch2020}, the design artifact can be parsed as an image, scored with a prediction from a Convolutional Neural Network, and explained with saliency maps \cite{Selvaraju2020} to identify problematic regions. For audio ideation tasks like music creation \cite{Louie2020}, Recurrent Neural Networks can be trained and explained with attention mechanisms \cite{Vaswani2017}. 

Interpretable Directed Diversity supports three popular explanation types, and future work can study others to stimulate creativity. For example, feature visualizations \cite{Olah2017} provide a ``vocabulary" of filters to interpret image predictions. For text, this could indicate concepts that ideators could reflect on to generate high scoring ideations; 
though these latent concepts may be uninterpretable. 
Concept Activation Vectors (CAVs) \cite{Kim2018} provide explanations with user-chosen concepts. Users could use CAVs like in SMILEY \cite{Cai2019} to discover ideations that are more similar to their desired concept.

\subsection{Human-AI Collaboration for Creativity}
Among the myriad techniques to support crowd ideation with human-guided feedback \cite{dow_shepherding_2012,Duque-Estrada2014,Oppenlaender2020a}, and automatic feedback \cite{Bae2020,Clark2018}, we add yet another technique to improve automation and scale. Similar to \cite{Louie2020}, our method directs or steers crowdworkers towards more creative ideas. Can human facilitators then leave the management to AI? No, we do not argue that crowd and feedback management be handled fully automatically. Instead, data management should be collaborative between the human facilitator and AI. Particularly, there is a need for better curation of source data for prompts and counterfactual suggestions (e.g., regularly update corpus \cite{Cox2021}). Human support is especially needed for the early phase, since there would be insufficient ideations to train an AI model to predict scores. This presents a causality dilemma that (ideation) labeling needs pre-collected, pre-labeled data for training, yet crowdsourcing is needed to label the initial data.

\section{Conclusion}
We have proposed Interpretable Directed Diversity to provide feedback to direct crowd ideators with explainable AI (XAI) feedback to iteratively generate more collectively creative ideas. We implemented and evaluated automatic scoring models and explanation techniques for Attributions, Contrastive Attributions, and Counterfactual Suggestions to improve ideation diversity and quality. Through a series of formative ideation study, summative ideation user study, and validation user studies, with computational language modeling embedding-based metrics and subjective user ratings, we found significantly positive effects of the proposed XAI types on increasing collective ideation diversity, except the Counterfactual Suggestions that still require improvements. Our results demonstrate the use of XAI for creativity applications. Hence, Interpretable Directed Diversity provides a generalizable method for scalable, real-time, and contextualized feedback to improve collective creativity.

\begin{acks}
This work was carried out in part at NUS Institute for Health Innovation and Technology (iHealthtech) and with funding support from the NUS ODPRT and Ministry of Education, Singapore.
\end{acks}

\bibliographystyle{ACM-Reference-Format}
\bibliography{DD2-YW}

\appendix

\onecolumn
\section{Appendix}

\subsection{Example training data for Quality (Motivation) machine learning model}

\newcolumntype{Z}{>{\centering\arraybackslash}X<{\hsize=.15\hsize}}

\begin{table*}[h]
\renewcommand{\arraystretch}{1.2}
\caption{Example messages from \cite{Cox2021} with varying motivation ratings.}
\vspace{-0.2cm}
 \label{table4}
 \small
\begin{tabularx}{\textwidth}{B cc}
\toprule
 \textbf{Message} & \textbf{Motivation\newline~Rating (–3 to 3)} & \textbf{Highly\newline~Motivating?} \\
 \midrule
 Walking outside in nature can rejuvenate your mind while toning your body at the same time! & 2.60 & High \\
 Push yourself, don’t wreck yourself. Set some steady goals, and work your way up! & 2.00 & High \\
 Even small amounts of exercise can help you become healthy. You don’t need to dedicate 2 hours a day to become healthier! & 1.75 & High\\
 Your dog would love to go on a walk with you each day! & 1.00 & Low\\
 Switch off an air conditioner while working out. Let the sweat out, and burn some calories. & 0.00 & Low \\
 Athletes are healthy since they train every day.  You will feel much healthier if you exercise too! & –0.17 & Low \\
\bottomrule
\end{tabularx}
\end{table*}

\clearpage

\subsection{Questionnaire Interfaces for User Studies}

\subsubsection{Summative Ideation User Study}
Screenshots for survey in the Feedback Condition SAXC, containing all explanation types. Other conditions will have fewer instructions and questions.

\begin{figure*}[h]
  \centering
  \includegraphics[width=3.8in]{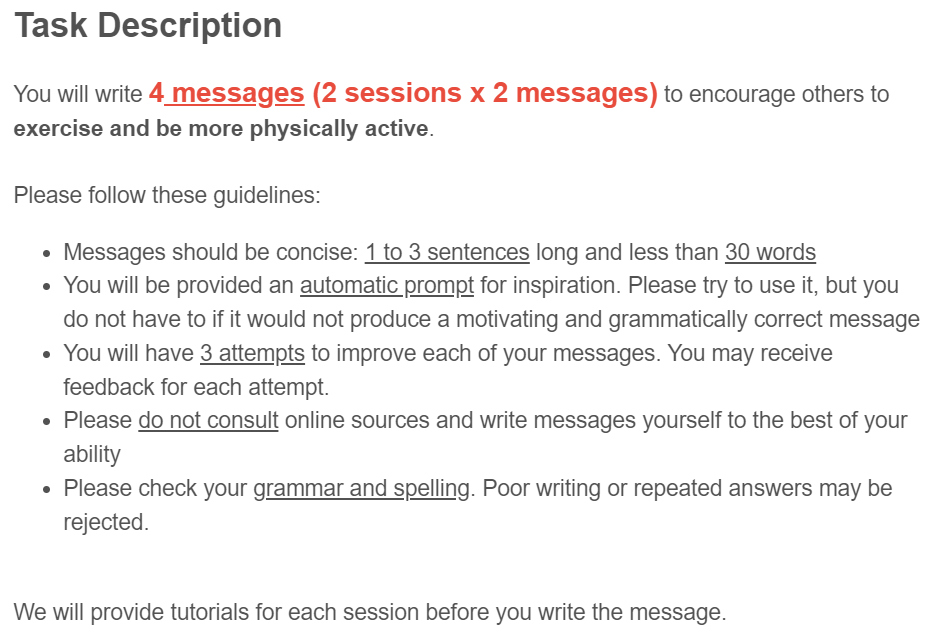}
  \caption{Introduction.}
  \Description{Screenshot of the Task Description in Summative Ideation User Study.}
  \label{fig14}
\end{figure*}

\begin{figure*}[h]
  \centering
  \includegraphics[width=5.4in]{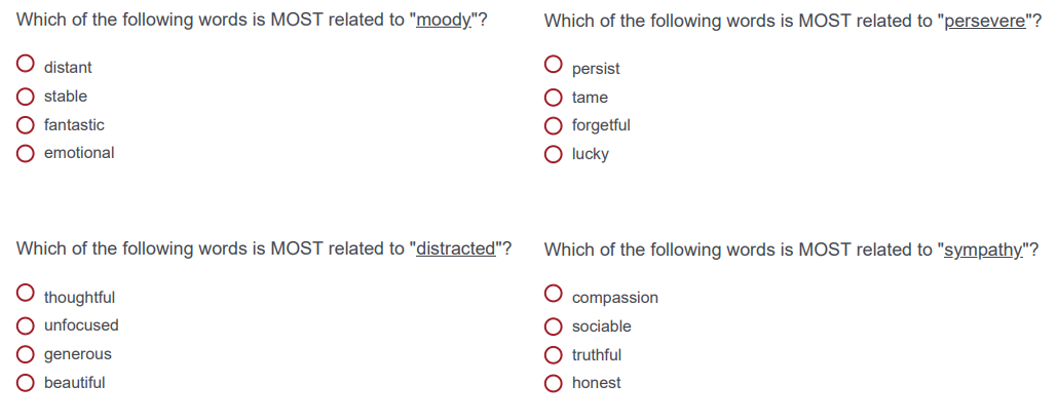}
  \caption{Screening questions on word association to assess English language skills.}
  \Description{Screenshot of the screening questions on word association to assess English language skills in Summative Ideation User Study.}
  \label{fig15}
\end{figure*}

\clearpage

\begin{figure*}[h]
  \centering
  \includegraphics[width=4.32in]{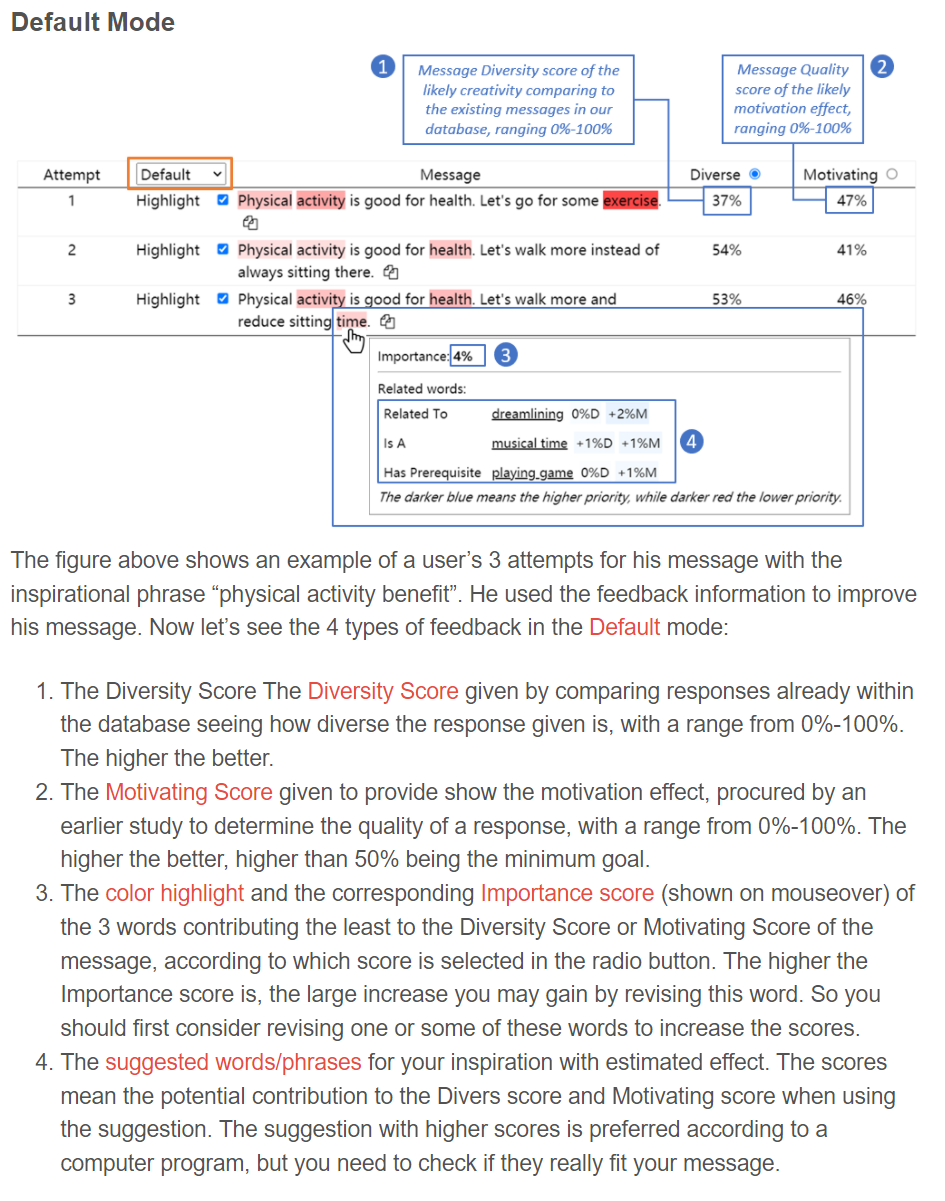}
  \caption{Tutorial on Scores, Attribution and Counterfactual explanations.}
  \Description{Screenshot of the tutorial on Scores, Attribution and Counterfactual explanations in Summative Ideation User Study.}
  \label{fig16}
\end{figure*}

\clearpage

\begin{figure*}[h]
  \centering
  \includegraphics[width=4.34in]{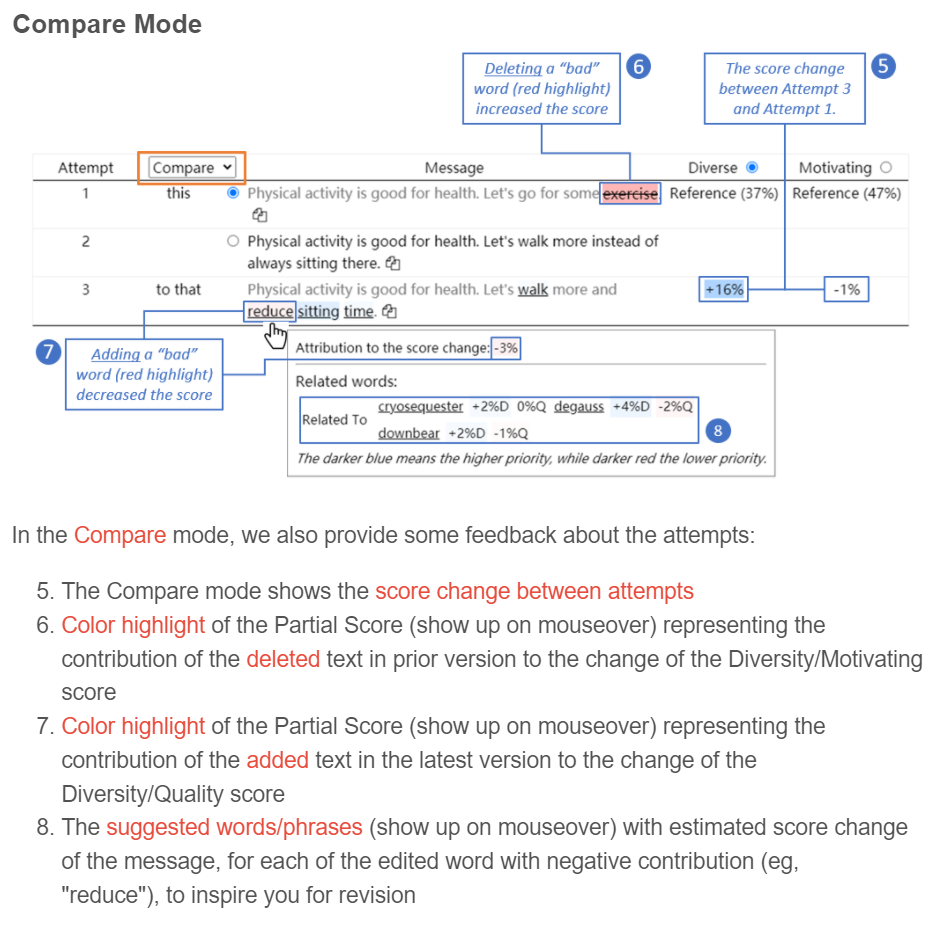}
  \caption{The tutorial of introducing the Contrastive explanations (i.e., color highlights and attribution to score changes) between attempts and Counterfactual explanations (i.e., suggested related words and potential contribution scores).}
  \Description{Screenshot of the tutorial on Contrastive explanations between attempts and Counterfactual explanations in Summative Ideation User Study.}
  \label{fig17}
\end{figure*}

\clearpage

\begin{figure*}[h]
  \centering
  \includegraphics[width=4.22in]{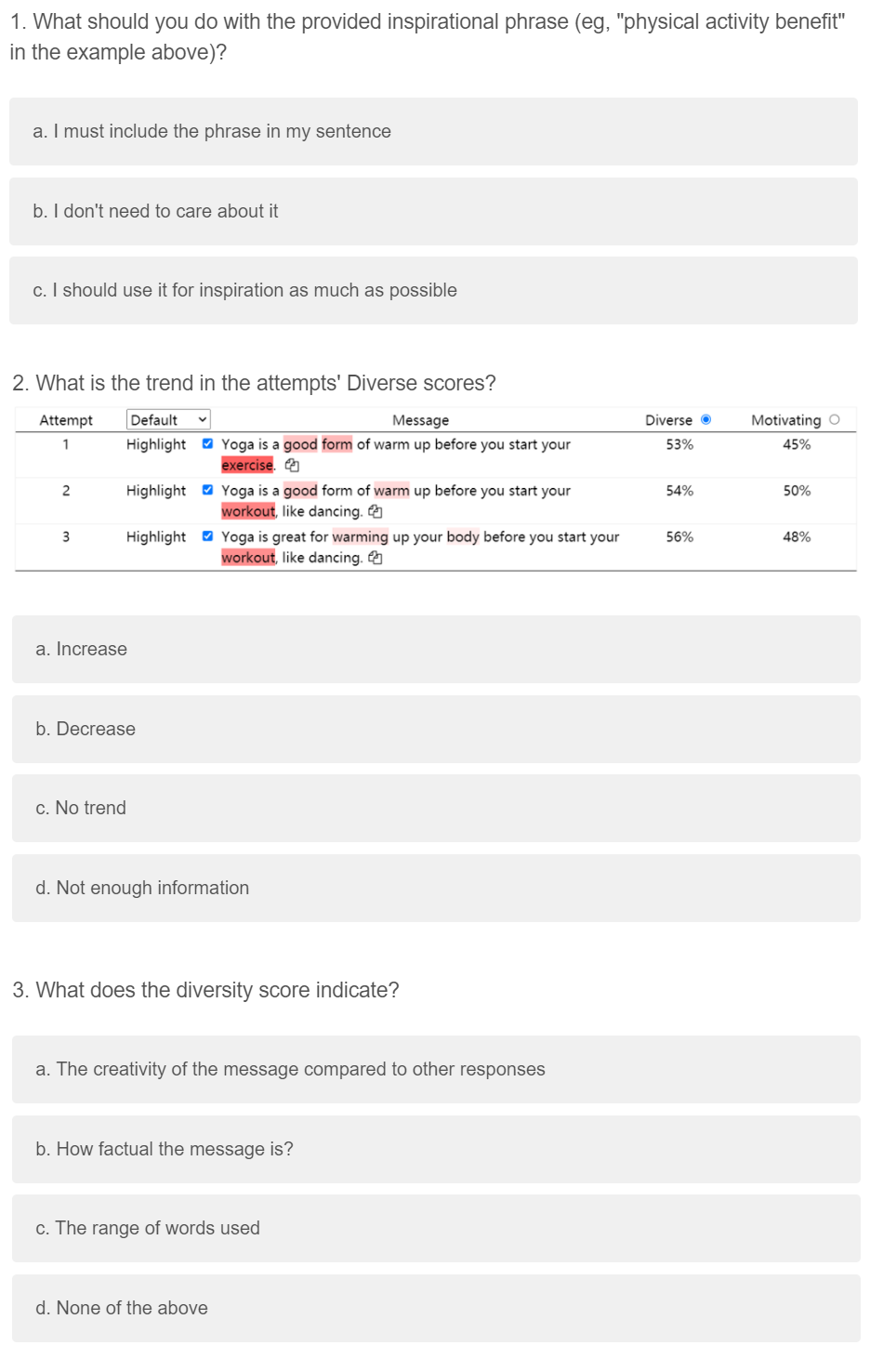}
  \caption{Screening Quiz.}
  \Description{Screenshot (part 1) of the screening quiz after the tutorial in Summative Ideation User Study.}
  \label{fig18a}
\end{figure*}

\clearpage

\setcounter{figure}{17}
\begin{figure*}[h]
  \centering
  \includegraphics[width=3.81in]{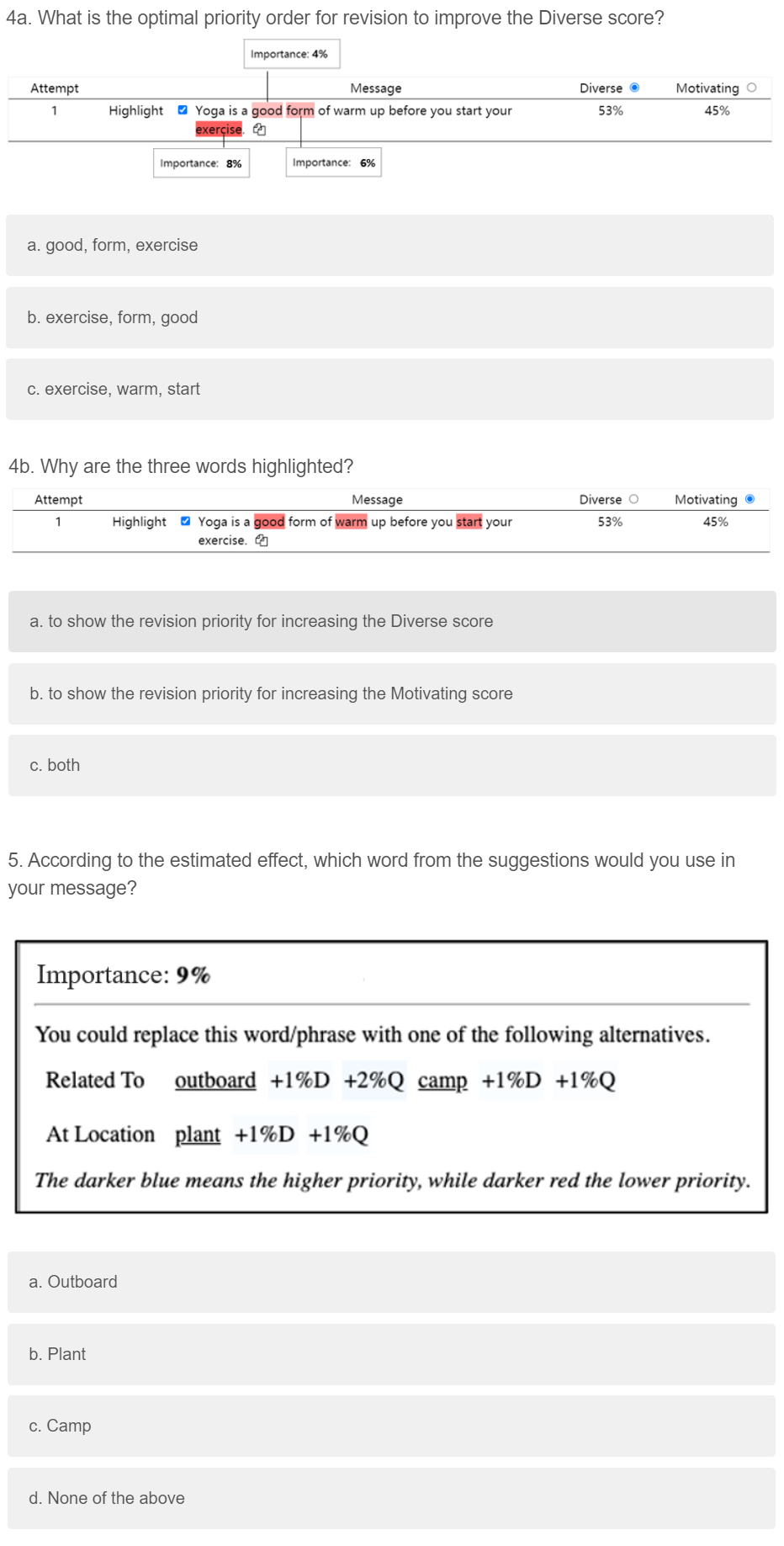}
  \caption{Screening Quiz. \textit{(continued)}}
  \Description{Screenshot (part 2) of the screening quiz after the tutorial in Summative Ideation User Study.}
  \label{fig18b}
\end{figure*}

\clearpage

\setcounter{figure}{17}
\begin{figure*}[h]
  \centering
  \includegraphics[width=4.25in]{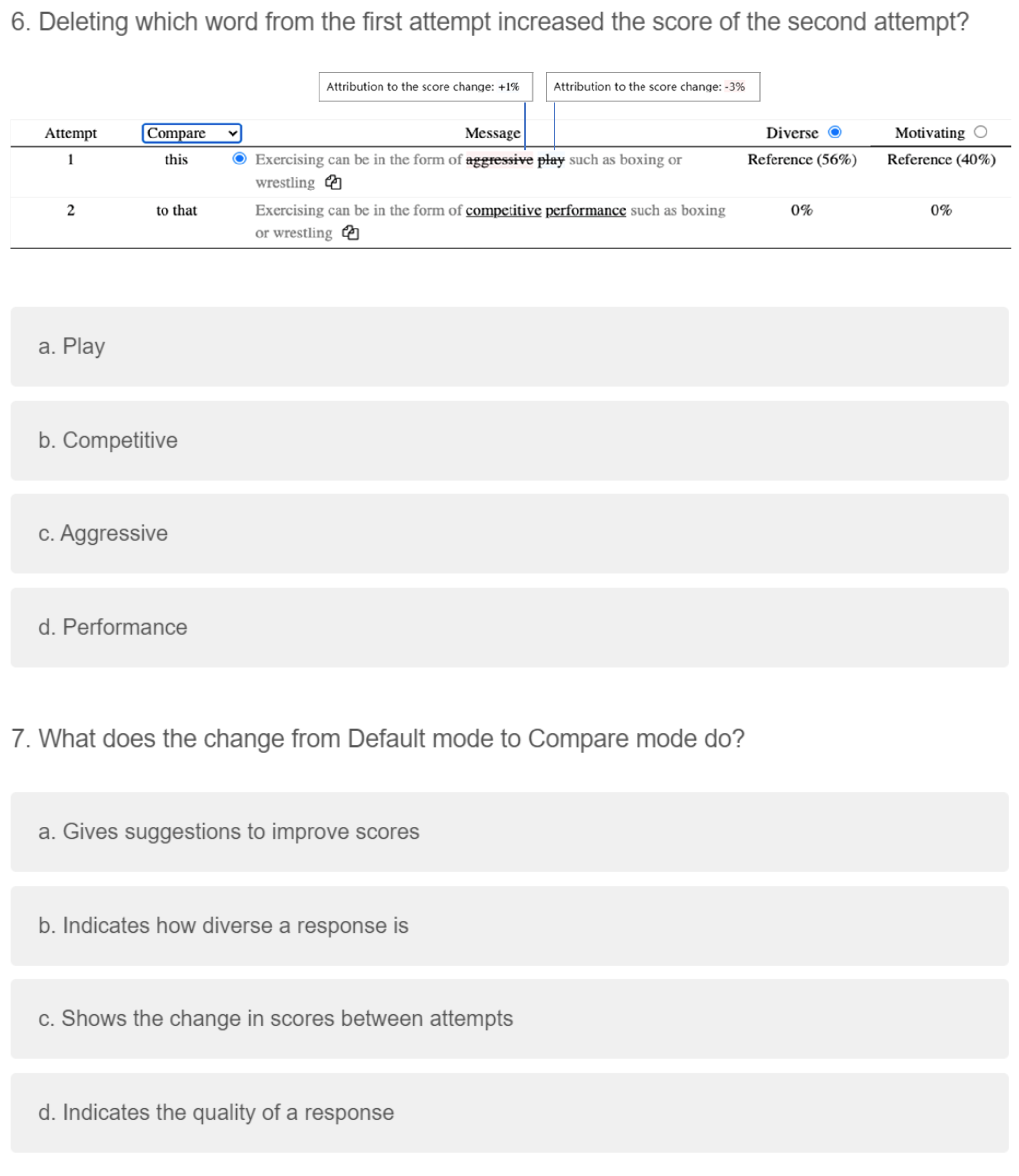}
  \caption{Screening Quiz. \textit{(continued)}}
  \Description{Screenshot (part 3) of the screening quiz after the tutorial in Summative Ideation User Study.}
  \label{fig18c}
\end{figure*}

\begin{figure*}[h]
  \centering
  \includegraphics[width=4.28in]{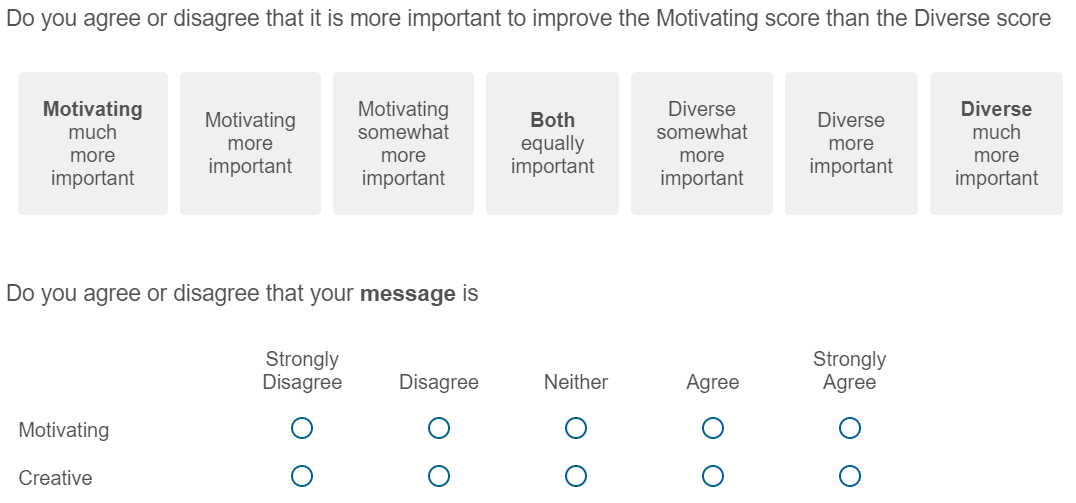}
  \caption{Questions to rate priority on Motivation or Creativity.}
  \Description{Screenshot of questions to rate priority on Motivation or Creativity after the ideation session in Summative Ideation User Study.}
  \label{fig19}
\end{figure*}

\clearpage

\begin{figure*}[h]
  \centering
  \includegraphics[width=3.91in]{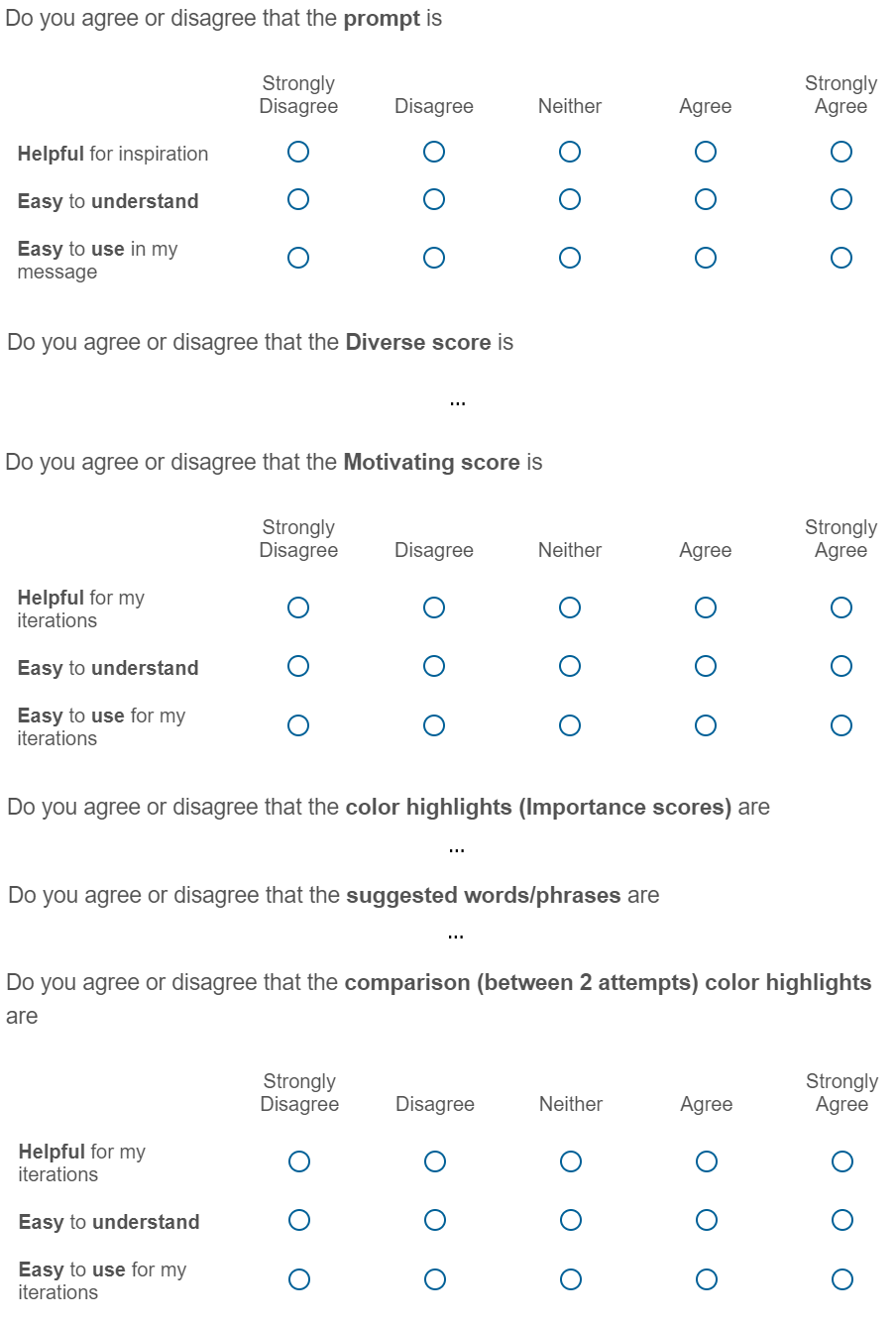}
  \caption{Questions on perceived helpfulness, ease of understanding, and ease of use regarding Feedback Features. Ellipses (…) notation indicate the same matrix of radio buttons, used for brevity.}
  \Description{Screenshot of questions to rate perceived helpfulness, ease of understanding, and ease of use regarding Feedback Features after the ideation session in Summative Ideation User Study.}
  \label{fig20}
\end{figure*}

\clearpage

\subsubsection{Ideation Quality Validation User Study} \hspace{0pt}

\begin{figure*}[h]
  \centering
  \includegraphics[width=3.9in]{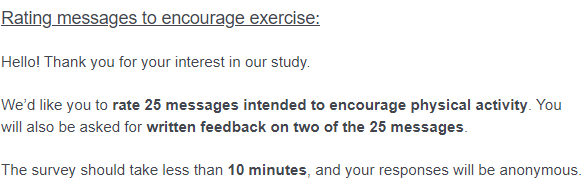}
  \caption{Introduction.}
  \Description{Screenshot of Task Description in Ideation Quality Validation User Study.}
  \label{fig21}
\end{figure*}

\begin{figure*}[h]
  \centering
  \includegraphics[width=4.78in]{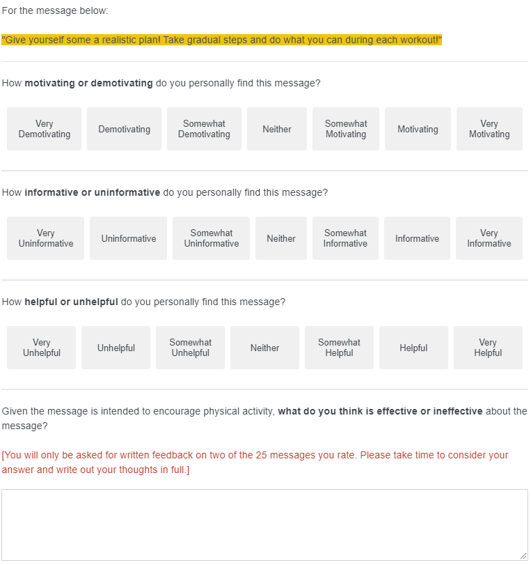}
  \caption{Questions to rate qualities of Motivatingness, Informativeness, and Helpfulness. Rationale was collected as attention checks.}
  \Description{Screenshot of questions to rate qualities of Motivatingness, Informativeness, and Helpfulness in Ideation Quality Validation User Study.}
  \label{fig22}
\end{figure*}

\clearpage

\subsubsection{Ideation Diversity Validation User Study} \hspace{0pt}

\begin{figure*}[h]
  \centering
  \includegraphics[width=3.91in]{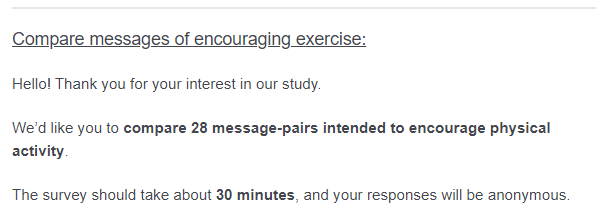}
  \caption{Introduction.}
  \Description{Screenshot of Task Description in Ideation Diversity Validation User Study.}
  \label{fig23}
\end{figure*}

\begin{figure*}[h]
  \centering
  \includegraphics[width=3.9in]{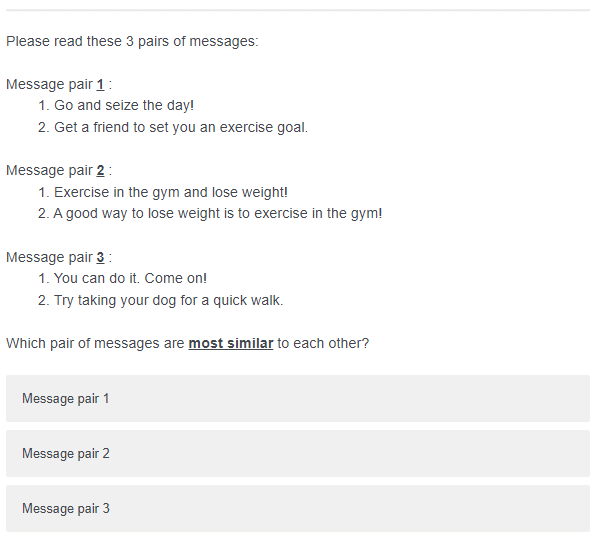}
  \caption{Screening quiz questions on message pair dissimilarity perception.}
  \Description{Screenshot of screening quiz questions on message pair dissimilarity perception in Ideation Quality Validation User Study.}
  \label{fig24}
\end{figure*}

\clearpage

\subsection{Linear Mixed Models and Statistical Analysis Results} \hspace{0pt}

\begin{table*}[h]
\caption{Statistical analysis for Ideation User Study of perceived ratings due to Feedback Feature as fixed effect and Participant as random effect in linear mixed effects models. \textcolor{lightgray}{\textit{n.s.}} means not significant at $p>.05$. $p>F$ is the significance level of the fixed effect ANOVA. $R^2$ is the model's coefficient of determination to indicate goodness of fit.}
\label{table5}
\begin{tabularx}{0.5\textwidth}{llrr}
 \toprule
 Response & 
 \shortstack{Linear Effects Model \\ (Participant as random effect)} & 
 $p>F$ & 
 $R^2$ \\
 \midrule
 Usefulness & Feedback Feature & <.0001 & .487\\
\grayline
 Ease of use & \textcolor{lightgray}{Feedback Feature} & \textcolor{lightgray}{\textit{n.s.}} & .480\\
\grayline
 Understandability & Feedback Feature & <.0001 & .471\\
 \bottomrule
\end{tabularx}
\end{table*}

\begin{table*}[h]
\caption{Statistical analysis for Ideation User Study of responses due to Feedback Condition, Prompt, and Score Increases as fixed effects and Participant as random effect in linear mixed effects models. $p>F$ is the significance level of the fixed effect ANOVA. $R^2$ is the model's coefficient of determination to indicate goodness of fit.}
\label{table6}
\begin{tabularx}{0.545\textwidth}{llrr}
 \toprule
 Response & 
\shortstack{Linear Effects Model \\ (Participant as random effect)} & 
 $p>F$ & 
 $R^2$ \\
 \midrule
 Ideation Time Effort & \textcolor{lightgray}{Feedback Condition +} & \textcolor{lightgray}{.0117} & .586\\
 & \textcolor{lightgray}{Prompt} & \textcolor{lightgray}{\textit{n.s.}} & \\
\grayline
 Quality Score Increase & \textcolor{lightgray}{Feedback Condition +} & \textcolor{lightgray}{.0361} & .384 \\
 & \textcolor{lightgray}{Prompt +} & \textcolor{lightgray}{\textit{n.s.}} & \\
 & Diversity Score Increase & <.0001 & \\
\grayline
 Diversity Score Increase & \textcolor{lightgray}{Feedback Condition +} & \textcolor{lightgray}{.0361} & .388 \\
 & \textcolor{lightgray}{Prompt +} & \textcolor{lightgray}{\textit{n.s.}} & \\
 & Quality Score Increase & <.0001 & \\
 \bottomrule
\end{tabularx}
\end{table*}
\begin{table*}[h]
\caption{Statistical analysis for Ideation User Study of computational diversity metrics due to Feedback Condition as fixed effect and Participant as random effect in linear mixed effects models. $p>F$ is the significance level of the fixed effect ANOVA. $R^2$ is the model's coefficient of determination to indicate goodness of fit.}
\label{table7}
\begin{tabularx}{0.585\textwidth}{llrr}
 \toprule
 Response & 
\shortstack{Linear Effects Model \\ (Participant as random effect)} & 
 $p>F$ & 
 $R^2$ \\
 \midrule
 \shortstack[l]{Ideation Dispersion\\ (MST Mean of Edge Weights)} & Feedback Condition & <.0001 & .405\\
\grayline
\shortstack[l]{Ideation Disparity\\ (Mean Pairwise Distance)} & Feedback Condition & <.0001 & .360\\
 \bottomrule
\end{tabularx}\vspace*{-6pt}
\end{table*}
\begin{table*}[h]
\caption{Statistical analysis for Ideation Quality Validation User study of responses due to Feedback Condition as fixed effects and Participant and Ideation as random effects in linear mixed effects models. $p>F$ is the significance level of the fixed effect ANOVA. $R^2$ is the model’s coefficient of determination to indicate goodness of fit.}
\label{table8}
\begin{tabularx}{0.61\textwidth}{llrr}
 \toprule
 Response & 
\shortstack{Linear Effects Model \\ (Participant, Ideation as random effects)} & 
 $p>F$ & 
 $R^2$ \\
 \midrule
 Motivatingness Rating & \textcolor{lightgray}{Feedback Condition} & \textcolor{lightgray}{\textit{n.s.}} & .292\\
\grayline
 Informativeness Rating & \textcolor{lightgray}{Feedback Condition} & \textcolor{lightgray}{\textit{n.s.}} & .298\\
\grayline
 Helpfulness Rating & \textcolor{lightgray}{Feedback Condition} & \textcolor{lightgray}{\textit{n.s.}} & .324\\
 \bottomrule
\end{tabularx}
\end{table*}

\begin{table*}[h]
\caption{Statistical analysis for Ideation Diversity Validation User study of responses due to Feedback Condition, Trial Index (sequence number of pair trial), \# Rationales Considered (how thoughtful participant was) and an interaction as fixed effects, and Participant as random effect in a linear mixed effects model. $p>F$ is the significance level of the fixed effect ANOVA. $R^2$ is the model's coefficient of determination to indicate goodness of fit.}
\label{table9}
\begin{tabularx}{0.71\textwidth}{llrr}
 \toprule
 Response & 
\shortstack{Linear Effects Model \\ (Participant as random effect)} & 
 $p>F$ & 
 $R^2$ \\
 \midrule
 Pairwise Dissimilarity Rating & Feedback Condition + & <.0001 & .371 \\
 & \textcolor{lightgray}{Trial Index +} & \textcolor{lightgray}{.0128} & \\
 & \textcolor{lightgray}{\# Rationales Considered +} &\textcolor{lightgray}{\textit{n.s.}} & \\
 & \textcolor{lightgray}{\# Rationales Considered × Feedback Condition} & \textcolor{lightgray}{\textit{n.s.}} & \\
 \bottomrule
\end{tabularx}
\end{table*}

\end{document}